\documentclass[a4paper,aps,pra,twocolumn,
tightenlines,preprintnumbersnofootinbib,showkeys,dvipsnames]{revtex4-1}
\usepackage{framed}
\usepackage[T1]{fontenc}
\usepackage{mathptmx}
\usepackage[framemethod=tikz]{mdframed}
\usepackage{bm}
\usepackage{bbold}
\usepackage{pifont}
\usepackage{rotating}
\usepackage{fullpage}
\usepackage{amsmath,amsthm,amssymb,mathrsfs}
\usepackage{slashed}
\usepackage{siunitx}
\usepackage{graphicx}
\usepackage[justification=centering]{caption}
\usepackage{amsthm}
\usepackage{epic}
\usepackage{eepic}
\usepackage{epsfig}
\usepackage{latexsym}
\usepackage[export]{adjustbox}
\usepackage{float}
\usepackage{multirow, makecell}
\usepackage{hyperref}
\usepackage{cleveref} 
\usepackage{enumitem}
\hypersetup{colorlinks=true,citecolor=red,linkcolor=NavyBlue,urlcolor=NavyBlue}
\usepackage[caption=false]{subfig}
 
\usepackage{natbib}
\usepackage{relsize}
\usepackage{footmisc}
\usepackage[left=1.6cm,right=1.6cm,top=2cm,bottom=2cm]{geometry}
\newtheorem{theorem}{Theorem}
\usepackage{tikz,lipsum,lmodern}
\usepackage{ulem} 
\usepackage{soul}
\usepackage{tcolorbox}
\usepackage{xcolor}
\usepackage[T1]{fontenc}
\restylefloat{figure}

\newcommand{\pwisein}{\left\{ \begin{array}{ll}}
\newcommand{\pwiseout}{\end{array}\right.}

\begin{document}
\title{Revocation and Reconstruction  of Shared Quantum States}
\author{Prakash Mudholkar $^1$}
\author{\rm Chiranjeevi Vanarasa $^1$}
\author{\rm Indranil Chakrabarty $^{1,2}$}
\author{\rm Srinathan Kannan $^{1,2}$}
\affiliation{$^1$ Center for Security, Theory and Algorithmic Research, International Institute of Information Technology, Gachibowli, Hyderabad 500 032, Telangana, India.}
\affiliation{ $^2$ Centre for Quantum Science and Technology,\\  International Institute of Information Technology, Gachibowli, Hyderabad 500 032, Telangana, India.}
\affiliation{\textit {\{prakash.mudholkar@research.iiit.ac.in, prakashvm@gmail.com, chiranjeevi.v@research.iiit.ac.in, indranil.chakrabarty@iiit.ac.in, srinathan@iiit.ac.in\}}}
\begin{abstract}
    The problem of revocation of quantum states after sharing is interesting and we ask: \textit{Is it possible for a dealer to revoke the state once shared, before the reconstruction process?} Additional resources like bell states are used to help the dealer to get back the state \cite {sazim2015retrieving}. In a three-party scenario, we show an independent way to revoke, if, for any reason, the dealer is not sure about the intention of the/any reconstructor.  In general, the classical outcomes of the dealer in sharing phase are needed, to be able to reconstruct the  state perfectly. When both the shareholders are dishonest\footnote{ By \textbf{semi-honest} share holders, we mean that the receivers diligently participate in the protocol, follow and execute the protocol as well, but are curious to know about the state of the dealer, and also desire to take other parties’ private information, but cannot collude with dishonest or malicious parties\cite{zhang2020three}. By \textbf{dishonest} share holders, we mean that the receivers may use fake shares in the  reconstruction phase\cite{harn2017share} or try to sabotage the entire action\cite{hillery1999quantum}.}, and without the dealer's knowledge, collude to reconstruct, they always have some chance of succeeding. This is addressed by giving more control to the dealer by making him/her to have a quantum share as well. We give a sharing and revocation protocol with a four-qubit entangled resource shared among three parties (two qubits with the dealer and one each with the shareholders). We further consider a class of four qubit pure entangled states as resource and explicitly find the range of parameters for which the protocol will be successful.
\end{abstract}
\maketitle    
\section{ Introduction}
\noindent In classical cryptography, secret sharing is a protocol where some sensitive classical data is distributed among number of people such that  sufficiently large subsets of those people can collectively reconstruct the data, but they can gain no information about the sensitive data from their respective individual shares. For instance, a possible application is to share the key for accessing a joint  account controlled by many people.  No individual is able to withdraw money, but sufficiently large groups can withdraw money or access the account. One particular symmetric variety of secret sharing scheme is called a $((k, n))$ threshold scheme, with $k \le n$ \cite{Blakley2011}, 
which allows any $k$ people out of $n$ to reconstruct the secret, while any set of up to $k - 1$ or fewer people has absolutely no information about that secret.\\ 

\noindent Over the past few decades we have seen rapid development of quantum information theory enabling us with various information processing tasks like, sending of quantum information \cite{bennett1993teleporting,sohail2023teleportation}, communication of classical information \cite{bennett1992communication,srivastava2019one}, key generation \cite{shor2000simple,ekert1991quantum}, remote state preparation \cite{bennett2005remote,pati2000minimum} and remote entanglement generation \cite{sazim2013study}, which are either better than the classical ways, or there are no classical means to carry out the tasks. Quantum secret sharing (QSS) \cite{hillery1999quantum,cleve1999share,singh2024controlled}, being one of such task, deals with the problem of sharing of both classical as well as quantum secrets. However, the resources used for secret sharing are purely quantum mechanical in nature. A typical protocol for quantum secret sharing, like many other tasks \cite{bennett1993teleporting,bennett1992communication,bennett2005remote,pati2023teleportation,wehner2018quantum,chakrabarty2010teleportation,chakrabarty2011deletion} uses quantum entanglement \cite{bell1964einstein}  as a cardinal resource, mostly pure entangled states. Karlsson et al.\cite{karlsson1999quantum} studied quantum secret sharing protocols using bipartite pure entangled states as resources. Many authors investigated the concept of quantum secret sharing  using tripartite pure entangled states and  multi-partite states like graph  states \cite{keet2010quantum,bandyopadhyay2000teleportation,li2009generation, markham2008graph,garg2024estimation,ma2025quantum,basak2025resource}. Q. Li et al. \cite{li2010semiquantum} proposed semi-quantum secret sharing protocols taking maximally entangled GHZ state as resource.
In a realistic situation however, the secret sharing of classical or quantum information involves transmission of qubits through noisy channels that give rise to mixed states. Several works were done in this context where mixed and noisy entangled states were taken as a resource for the purpose of secret sharing \cite{adhikari2010probabilistic,ray2016sequential}. 
Quantum secret sharing has also been realized in experiments \cite{tittel2001experimental,schmid2005experimental,schmid2006experimental,bogdanski2008experimental,yan2025quantum,zhang2025high}, and various reconstruction circuits for quantum secret sharing are proposed \cite{chiwaki2025measurement}, and the security aspect was investigated separately \cite{grilo2025security}. Earlier, quantum  state reconstruction and secret sharing are considered to be same until recently  a more stricter definition of quantum secret sharing is introduced \cite{singh2024controlled} and subsequently the class of states for which the maximal secret reconstruction is possible has been identified \cite{abrol2024secret}.  The old idea of sharing the information between the parties and then reconstructing the state at one location with the help of the party at the other location does not care about the limits of accessible information. It only ensures each party do not have full information and cannot access the secret independently. It is just referred to as controlled state reconstruction (CSR) in the reference \cite{singh2024controlled}, and in this article, we will be adapting the same terminology.\\
\noindent In section II, we build up the motivation for revoking the quantum state back after the sharing process, and before the reconstruction process. In section III, we explore the advantage of retaining a \textbf{quantum share}, by the dealer, and introduce our revocation and reconstruction protocol for one dealer and two share holders. We also provide an example of a resource state that is useful for our protocol. In section IV, we execute our protocol with a resource belonging to a class of four qubit pure entangled states \cite{verstraete2002four}. In particular, we find out the range of the parameters for the resource state, with which both, revocation at the dealer's location and the reconstruction at the share holder's location are possible. \\

\section{Motivation}

\noindent The entire idea of secret sharing can be understood broadly from two different perspective. One is the secret to be shared and another is resource to be used for sharing. Each of them can be either classical or quantum. Before the advent of quantum secret sharing in the standard classical secret sharing problem, the usual scenario is that there is a dealer who possesses a secret information (classical bits), and some other parties (share holders), with whom the secret information will be shared by the dealer during the sharing phase. In the reconstruction phase, parties as required by the protocol, with no help from the dealer, work together to reconstruct the secret information perfectly. Here in such a scenario there is no such quantum channel between the dealer and share holders. 

\noindent In quantum secret sharing there can be two possibilities where we are sharing classical bits or quantum bits, however the resource state through which sharing will be possible are always quantum. In sharing of quantum states  or sometimes known as splitting of quantum information\cite{hillery1999quantum}, at the end of the sharing phase, the dealer also has some information, i.e., two classical bits representing the outcome of the measurement done in sharing phase, which is essential to the shareholders in the reconstruction phase. 

\noindent It is important to mention here that the secret sharing of classical information is different from the sharing of the quantum information. In case of secret sharing of classical information, if both the receivers are dishonest, it is a no-win scenario as they can always clone their classical information on their part and combine to get back the original secret without hampering the original protocol. Now since here the state to be shared is a quantum state, the no-cloning theorem \cite{wootters1982single} does not allow at the very first place to clone the state by the shareholders. So, it becomes necessary to investigate from the dealer’s point of view how to revoke the state if all the shareholders turn out to be dishonest.

\noindent Prior to the newly added concept of security in the reference \cite{singh2024controlled}, in most of the quantum state sharing scenarios, protocols ensure security only if some of the share holders are semi-honest. 
However these protocols fail to address the situation when the dealer after sharing finds all the share holders to be dishonest. Now let us compare the two situations, (i) where dealer only has the information about outcome of measurement in sharing phase, (ii) where dealer is also being one of the share holders, i.e., dealer himself has a quantum share.
In case (i), if share holders turns out to be dishonest and plan to collude, guessing the dealer's measurement outcome, to reconstruct the secret, they can succeed with non-zero probability (precisely 1/4 in Hillery et al's protocol\cite{hillery1999quantum}), where as in case (ii), as dealer is also a shareholder, she can delay her measurement to the last among all the measurements to be performed in the reconstruction phase so that it is impossible to collude and construct the secret perfectly ignoring the dealer's quantum share. In reference\cite{sazim2015retrieving}, a revocation proposal was introduced with the help of an additional Bell state between the dealer and the reconstructor. This requires a two qubit resource in addition to the shared three qubit GHZ state (which means that to perform the revocation, we need a total of five qubits). In the CSR protocol which we are proposing, we gain advantage over sabotaging (by the dishonest shareholders), by considering a quantum share, with the dealer, using a four qubit state. This helps to revoke the state back with resource having lesser number of  qubits as compared to five qubits in the protocol proposed by Sazim et. al. \cite{sazim2015retrieving}).

\noindent The protocol which we propose, primarily \textit{shields the state from dishonest players to get reconstructed}, even though it cannot stop them from sabotaging/destroying the secret, as it is a general situation in quantum protocols. 
Our protocol is based on a simple scenario when there is one dealer Alice and the state is shared between Bob and Charlie. Here in our protocol we empower the dealer to keep a part of the secret quantum state (which we refer to as quantum share) with herself which enables her to stop the reconstruction of the state by Bob and Charlie. Consequently, we also give a revocation protocol, when executed by all three parties can result in the dealer retrieving back the secret state. This secret sharing scheme shields the secret quantum state from dishonest adversaries, simultaneously accommodating revocation by the dealer, when the share holders are semi-honest. This empowerment comes at the cost of one extra qubit with the dealer. So for a three party scenario (one dealer and two additional share holders) we need a four qubit entangled state as a resource for the execution of the protocol.

\noindent The CSR protocol, which we propose, can be substituted wherever quantum secret sharing can be used, provided the required resource can be prepared and established among the involved parties. 
Quantum secret sharing schemes can be used 
in a multi-partite scenario in the quantum money\cite{10.1145/1008908.1008920, Gottesman_2000}, 
Quantum Networking, and a secure distributed quantum computation\cite{Gottesman_2000}. It can also be used in the context of sharing a quantum key securely\cite{refId0}. In this context, if all the share holders are dishonest, and one has to abort the process, then our protocol becomes useful as the dealer can revoke back the secret.

\section{The Revocation and Reconstruction Protocol: Quantum Share Of The State}
In this section, we first discuss the difficulty in carrying out this protocol in a three party, three qubit scenario. Further, we give an outline of our protocol in general and introduce the concept of \textbf{quantum share} of state in that process. In addition to that, we consider a four qubit state and demonstrate our protocol by taking that as resource.
\\

\subsection{Going Beyond Three Qubit State Sharing}
\noindent In most of the existing CSR protocols, the dealer (Alice), who has the state to be shared, will first split the quantum information by combining her state with the resource state, and then does a two qubit measurement on the to be shared-qubit and the part of the resource qubit shared with her. In this case Alice, Bob and Charlie (Bob and Charlie are the two parties with whom Alice wants to share the state) are sharing a three qubit resource among them with each possessing one qubit. 
The measurement will result in the unknown qubit being shared  with the other two parties, Bob and Charlie. Alice will only have classical outcome of the measurement but not a quantum part of the state that she has shared. The strength of 3-party, 3-qubit protocol is such that if one of the share holders (Bob or Charlie) is dishonest,
he can not reveal the state, as the honest
party will prevent the dishonest party from doing any damage\cite{hillery1999quantum}.  
However if both of them are dishonest, as given in 
\cref{tab:first_table},
they can collaborate with each other to get hold of the shared qubit, and Alice can not do anything about it.  This can allow Bob and Charlie, if dishonest,
to collude and reconstruct the exact  qubit with a probability of $1/4$.
\begin{table}
\begin{center}
\begin{tabular}{|c|c|c|}
\hline \parbox{1in}{\ \\\begin{center}Alice\end{center} \ \\} & \parbox{1in}{\ \\\begin{center}Bob\end{center} \ \\} & \parbox{1in}{\ \\\begin{center}Charlie\end{center}\  \ \\ }\\
\hline Dealer & Honest & Honest \\
\hline Dealer & Honest & Semi-Honest   \\
\hline Dealer & Semi-Honest & Honest   \\
\hline Dealer & Semi-Honest & Semi-Honest   \\
\hline Dealer & Honest & Dishonest   \\
\hline Dealer & Dishonest & Honest   \\
\hline Dealer & Semi-honest & Dishonest   \\
\hline Dealer & Dishonest & Semi-Honest   \\
\hline Dealer & Dishonest & Dishonest   \\
\hline
\end{tabular} 
\end{center}
\caption{Adversarial Nature of Parties Involved}
\label{tab:first_table}
\end{table}
\noindent Alice combines her qubit (state), $\alpha|0\rangle_A + \beta|1\rangle_A$, which she intends to share with Bob and Charlie, with her share in the 3-qubit GHZ resource $a, b, c$ such that only Alice has access to $a$, only Bob has access to $b$, and only Charlie has access to $c$ \cite{hillery1999quantum}. After measuring the pair in the Bell basis, she now conveys the result of her measurement to Bob or Charlie. 
Assuming Bob now measures his state in the Hadamard basis, Charlie is able to reconstruct the state. However he needs two bits of classical information in order to reconstruct it. Hence, after Alice now conveys the result of her measurement in Bell basis to Bob, Charlie can reconstruct Alice’s state but only with Bob's assistance\cite{hillery1999quantum}. Bob must measure his state and send the result to Charlie. Without Bob’s information, Charlie can not reconstruct. If Alice comes to know that  both are dishonest, 
she does not share her results of measurement in the Bell basis with Bob or Charlie. Even then, Bob and Charlie can reconstruct the exact state without the help of Alice with a probability of $1/4$. However, if Alice comes to know about the information that both are dishonest 
after conveying the results of her measurement in the Bell basis, there is nothing she can do to prevent Bob and Charlie from reconstructing the state. \\
\\
To nullify this probability, we introduce the need of one more qubit with Alice, so that she can also hold a share in the quantum form. This we refer as \textbf{quantum share} of the state. Thus, Bob and Charlie will be unable to reconstruct the state without the involvement of Alice. 
If Alice finds both the parties to be dishonest,
then she can always bring back the state to her, with the help of Bob and Charlie. 
\\ 
\subsection{Our Protocol: A New Quantum State Sharing Protocol}
%
%
%
\noindent In this subsection we give an outline of our protocol which requires a 4-qubit resource, with two of the qubits with Alice, the third qubit with Bob and the fourth qubit with Charlie.  \\
 
\begin{figure*}
\begin{framed}
          \begin{minipage}{1\textwidth}
            \centering
            \includegraphics[width=0.4\textwidth]{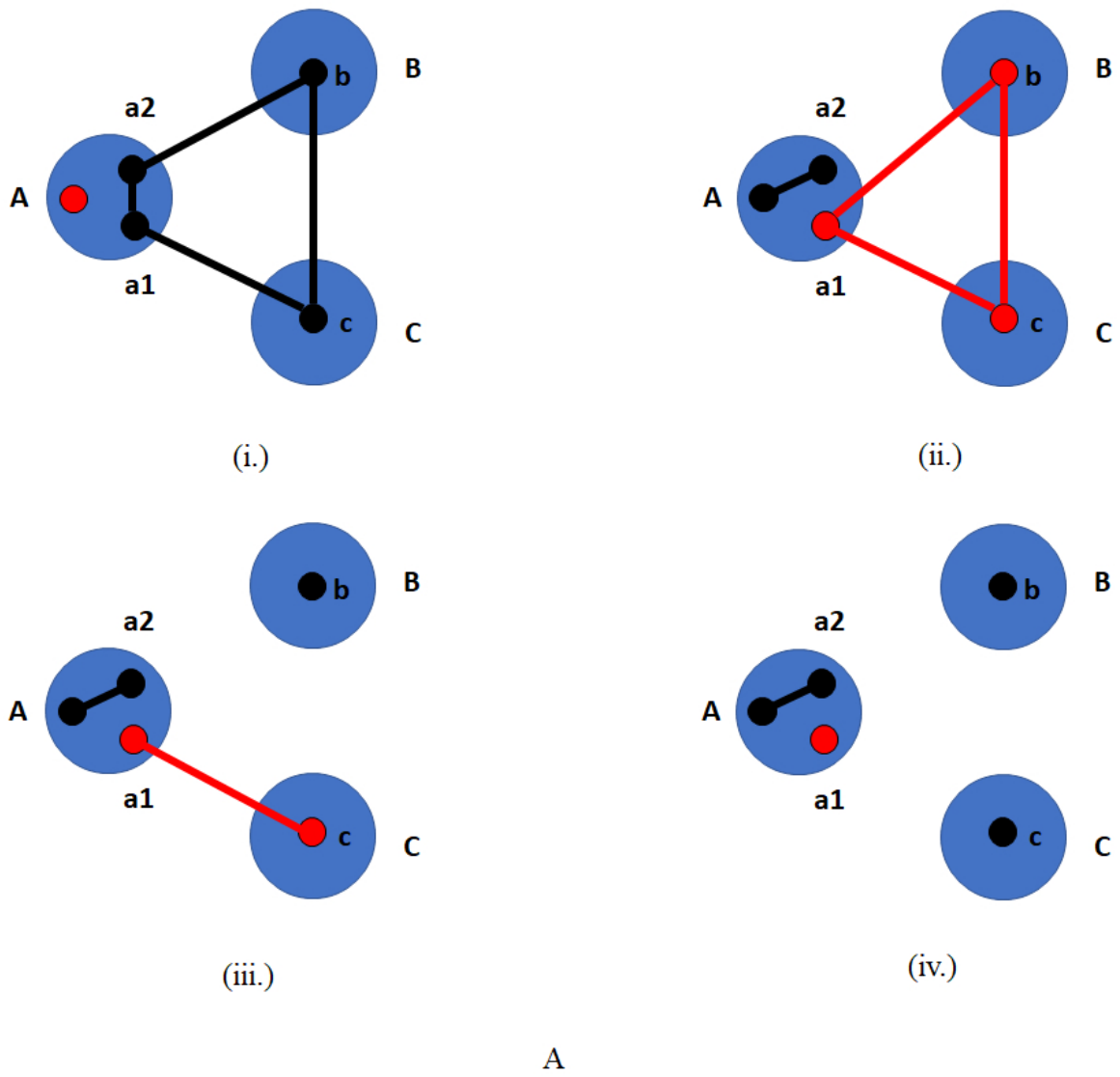}
            \includegraphics[width=0.4\textwidth]{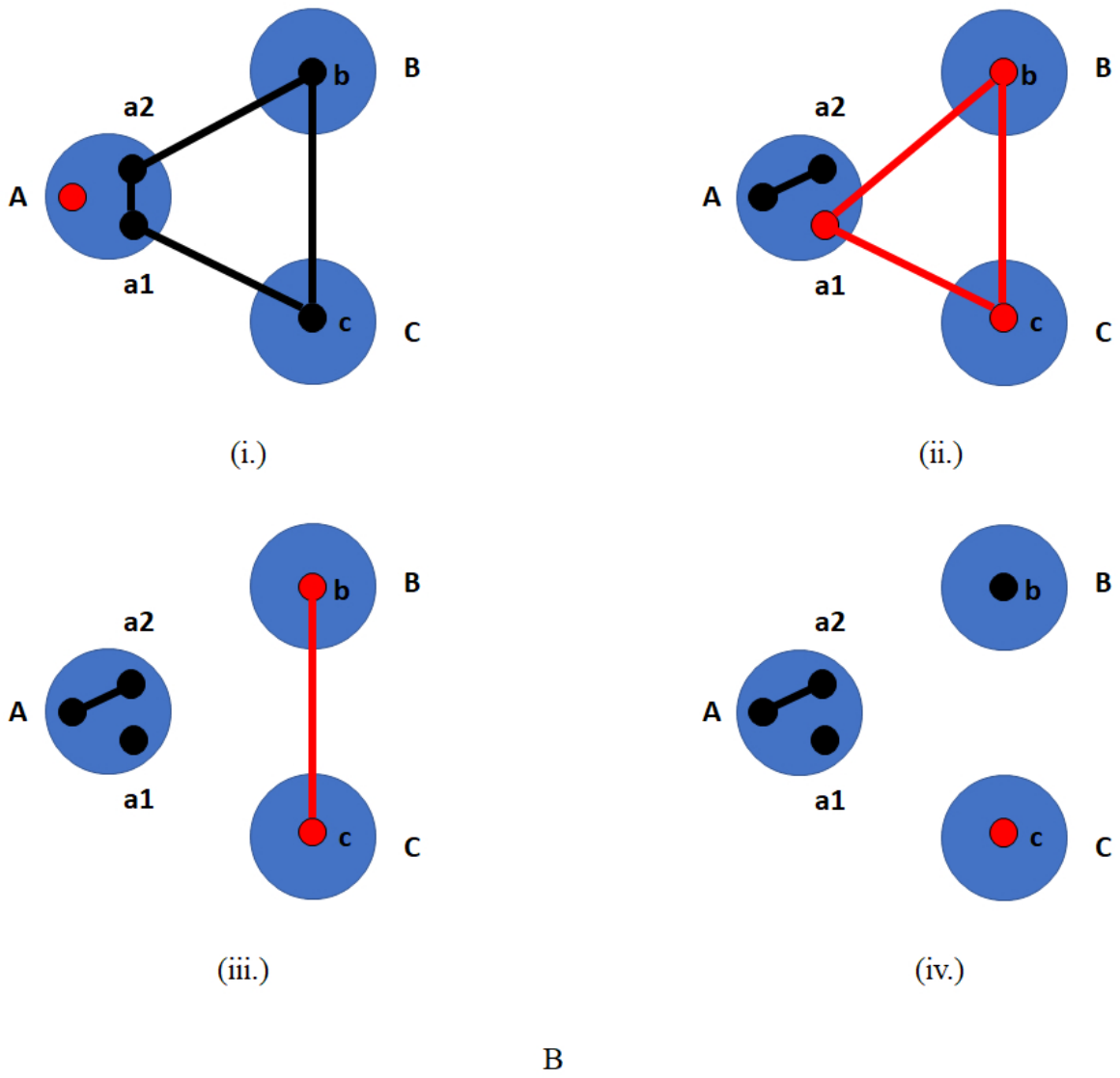}
          \end{minipage}
          \caption{$A(i.)$ Alice with the secret state (depicted in red), Alice, Bob and Charlie with the 4 qubit resource (depicted in black). $A(ii.)$ Three qubit entanglement among Alice, Bob and Charlie obtained after splitting of quantum information. $A(iii.)$ Two qubit entanglement between Alice and Charlie after Bob's measurement. $A(iv.)$ Revocation by Alice. $B(i.)$ Alice with the secret state (depicted in red), Alice, Bob and Charlie with the 4 qubit resource (depicted in black). $B(ii.)$ Three qubit entanglement among Alice, Bob and Charlie obtained after splitting of quantum information. $B(iii.)$ Two qubit entanglement between Bob and Charlie after Alice's measurement. $B(iv.)$ Charlie with the Reconstructed State.} 
          \label{fig:recon_figure}
          \end{framed}
        \end{figure*}

\noindent \textbf{Step 1: Splitting of Quantum Information or the sharing  phase.}
\noindent Alice combines the state $|\psi\rangle_{s}=\alpha|0\rangle_s + \beta|1\rangle_s$, (where $|\alpha|^2+|\beta|^2=1$) with the 4-Qubit entangled resources $|X\rangle_{abcd}$, where the qubits $a$ and $b$ are with Alice, while $c$ and $d$ are with Bob and Charlie respectively. She measures the state $s$ and the qubit $a$  in the Bell basis, 
\begin{eqnarray}
   |\phi^{\pm}\rangle = \frac{1}{\sqrt{2}}(|00\rangle \pm |11\rangle) \nonumber \\
    |\psi^{\pm}\rangle = \frac{1}{\sqrt{2}}(|01\rangle \pm |10\rangle) 
\end{eqnarray}
and she gets one of the four possible outcomes $\{\psi^{\pm}, \phi^{\pm}\}$ (Bell states). Corresponding to each measurement outcomes, Alice, Bob and Charlie share three qubit entanglement among each other. The state is shared in form of three qubit entanglement among these three parties .\\ 
\\
\noindent \textbf{Step 2: Revocation and Reconstruction of the state.}
\noindent Alice then uses her second qubit (\textbf{quantum share of the resource}) to participate in the reconstruction of the state along with Bob or Charlie. If Charlie needs to reconstruct the state, Alice and Bob measure their respective particles in the Hadamard basis (see \cref{fig:recon_figure}). At this point, if Alice comes to know that 
both, Bob and Charlie are dishonest, 
she will not measure her state and the chance that Charlie retrieves the state is zero. In addition, Alice can bring back her state by asking Bob and Charlie (though dishonest, but still willing to participate in the protocol) to do their measurements in the Hadamard basis. The importance of the protocol is that the dealer Alice holds a quantum share of the state with herself, which empowers her to stop the revealing of the state,  and consequently bring the entire state back to herself without revealing it even after sharing.
\\

\noindent \textbf{Example: Demonstration of our protocol with a four qubit state :}
\noindent Here we consider a 4-qubit normalized state to be used as an example to demonstrate our protocol for retrieving and reconstructing the state. The state is given as,
\begin{eqnarray}
 |G_{abcd}^E\rangle = \frac{1}{\sqrt{2}}(|0000\rangle+|1111\rangle) ,
\end{eqnarray}
where $E$ in $|G_{abcd}^E\rangle$ depicts a special case of the $|G_{abcd}\rangle$ resource state.
This is obtained by putting $a = \frac{1}{\sqrt{2}}, b = 0, c = 0, d = \frac{1}{\sqrt{2}}$ into  the state, 
\begin{eqnarray}
|G_{abcd}\rangle = &&\frac{a+d}{2}(|0000\rangle+|1111\rangle) + \frac{a-d}{2}(|1100\rangle+|0011\rangle){}\nonumber \\&& +\frac{b+c}{2}(|0101\rangle+|1010\rangle) +\frac{b-c}{2}(|0110\rangle+|1001\rangle){}\nonumber \\&&
\end{eqnarray}
Here $a$, $b$, $c$ and $d$ being complex numbers and normalization condition being $|a|^2+|b|^2+|c|^2+|d|^2 = 1$. In the state, the first two qubits are with Alice, the third one with Bob and the last one with Charlie and let $|\psi_{S}\rangle = \alpha|0\rangle+\beta|1\rangle$ be the state to be shared by Alice with Bob and Charlie.
\begin{eqnarray}
&&|\psi_{S}\rangle\otimes |G_{abcd}^E\rangle 
=\{\alpha|0\rangle+\beta|1\rangle\}\otimes \frac{1}{\sqrt{2}}\{(|0000\rangle+|1111\rangle){}\nonumber\\&&
=\frac{1}{\sqrt{2}}\{\alpha|00000\rangle+\alpha|01111\rangle 
+\beta|10000\rangle+\beta|11111\rangle\} 
\end{eqnarray}
\noindent On further simplification of the combined state,
\begin{eqnarray}
&&|\psi_{S}\rangle\otimes |G_{abcd}^E\rangle=(\frac{1}{2})\{(|\phi^+\rangle)[(\alpha|000\rangle+\beta|111\rangle)]{}\nonumber\\&& 
+(|\phi^-\rangle)[(\alpha|000\rangle-\beta|111\rangle)] +(|\psi^+\rangle)[(\alpha|111\rangle+\beta|000\rangle)]{}\nonumber\\&& +(|\psi^-\rangle)[(\alpha|111\rangle-\beta|000\rangle)]\} 
\end{eqnarray}
\noindent Let us consider a particular case when Alice measures in $|\phi^+\rangle$ Bell Basis. So the state already being shared between Alice, Bob and Charlie is:$|Y_{234}^{\phi^+}\rangle=(\frac{1}{2})\{(\alpha|000\rangle+\beta|111\rangle)\}$  
If we write Bob and Charlie's qubit in Hadamard Basis, then the combined shared system can be rewritten as,  
\begin{eqnarray}
&&|Y_{234}^{\phi^+}\rangle=(\frac{1}{4})\{[\alpha|0\rangle_{A}[(|+\rangle+|-\rangle)_{B} (|+\rangle+|-\rangle)_{C}]{}\nonumber\\&& +\beta|1\rangle_{A}[(|+\rangle-|-\rangle)_{B} (|+\rangle-|-\rangle)_{C}]]\} 
\end{eqnarray} 
Now, if Bob and Charlie measure their respective qubits in Hadamard Basis, Alice is able to retrieve the state,
$|Y_{234}^{\phi^+}\rangle = (\frac{1}{2})\{(\alpha|0\rangle+\beta|1\rangle)|+\rangle_{B}|+\rangle_{C}+(\alpha|0\rangle-\beta|1\rangle)|+\rangle_{B}|-\rangle_{C}+(\alpha|0\rangle-\beta|1\rangle)|-\rangle_{B}|+\rangle_{C}-(\alpha|1\rangle+\beta|0\rangle)|-\rangle_{B}|-\rangle_{C}\}$\\ 

\noindent \textbf{Case 1:} When Bob and Charlie measure in the Hadamard Basis $|+\rangle|+\rangle$, Alice's resultant state is as follows and the transformation which Alice should perform in order to retrieve/revoke the state, up to an overall
sign, is:
$(\alpha|0\rangle+\beta|1\rangle) \rightarrow I $\\

\noindent \textbf{Case 2:} When Bob and Charlie measure in the Hadamard Basis $|+\rangle|-\rangle$, Alice's resultant state is as follows and the transformation which Alice should perform in order to retrieve/revoke the state, up to an overall
sign, is:
$(\alpha|0\rangle-\beta|1\rangle) \rightarrow \sigma_z $\\

\noindent \textbf{Case 3:} When Bob and Charlie measure in the Hadamard Basis $|-\rangle|+\rangle$, Alice's resultant state is as follows and the transformation which Alice should perform in order to retrieve/revoke the state, up to an overall
sign, is:
$(\alpha|0\rangle-\beta|1\rangle) \rightarrow \sigma_z $\\ 

\noindent \textbf{Case 4:} When Bob and Charlie measure in the Hadamard Basis $|-\rangle|-\rangle$, Alice's resultant state is as follows and the transformation which Alice should perform in order to retrieve/revoke the state, up to an overall
sign, is:
$(\alpha|0\rangle+\beta|1\rangle) \rightarrow I$\\

\noindent {Now, if Alice and Bob measure their respective qubits in Hadamard Basis, Charlie (without any loss of generality we have taken Charlie to be the reconstructor) is able to reconstruct the state,}
$|Y_{234}^{\phi^+}\rangle = (\frac{1}{2})\{(\alpha|0\rangle+\beta|1\rangle)|+\rangle_{A}|+\rangle_{B}+(\alpha|0\rangle-\beta|1\rangle)|+\rangle_{A}|-\rangle_{B}+(\alpha|0\rangle-\beta|1\rangle)|-\rangle_{A}|+\rangle_{B}-(\alpha|1\rangle+\beta|0\rangle)|-\rangle_{A}|-\rangle_{B}\}$\\

\noindent \textbf{Case 1:} When Alice and Bob measure in the Hadamard Basis $|+\rangle|+\rangle$, Charlie's resultant state is as follows and the transformation which Charlie should perform in order to obtain Alice’s qubit, up to an overall
sign, is:
$(\alpha|0\rangle+\beta|1\rangle) \rightarrow I $\\

\noindent \textbf{Case 2:}  When Alice and Bob measure in the Hadamard Basis $|+\rangle|-\rangle$, Charlie's resultant state is as follows and the transformation which Charlie should perform in order to obtain Alice’s qubit, up to an overall
sign, is:
$(\alpha|0\rangle-\beta|1\rangle) \rightarrow \sigma_z $\\

\noindent \textbf{Case 3:} When Alice and Bob measure in the Hadamard Basis $|-\rangle|+\rangle$, Charlie's resultant state is as follows and the transformation which Charlie should perform in order to obtain Alice’s qubit, up to an overall
sign, is:
$(\alpha|0\rangle-\beta|1\rangle) \rightarrow \sigma_z $\\

\noindent \textbf{Case 4:} When Alice and Bob measure in the Hadamard Basis $|-\rangle|-\rangle$, Charlie's resultant state is as follows and the transformation which Charlie should perform in order to obtain Alice’s qubit, up to an overall sign, is:
$(\alpha|0\rangle+\beta|1\rangle) \rightarrow I$ $\blacksquare$\\
\noindent In \cref{tab:third_table}, 
we show how Alice can retrieve/revoke back her message by enlisting down the respective local operations corresponding to Bob’s and Charlie’s measurement outcomes when ($b = 0$ and $c = 0$) and ($
(a = \frac{1}{\sqrt{2}}$ and $d = \frac{1}{\sqrt{2}})$. In \cref{tab:third_table}, 
we also show the Reconstructing Quantum Information of Charlie’s resultant state by enlisting down the respective local operations corresponding to Alice's and Bob’s measurement outcomes when ($b = 0$ and $c = 0$) and ($
(a = \frac{1}{\sqrt{2}}$ and $d = \frac{1}{\sqrt{2}})$. Thus we are able to demonstrate our protocol for the four qubit state $|G_{abcd}^E\rangle$.
\vspace{-0.3in}
\\ \\
\section{ Characterization of Four Qubit States  useful for Protocol} 
\noindent In this section we consider a particular class of four qubit state \cite{verstraete2002four} and investigate whether they will be useful to be used as resource states for our protocol. We begin with the first $4-$qubit normalized $G_{abcd}$ state:
\begin{eqnarray}
|G_{abcd}\rangle & = &\frac{a+d}{2}(|0000\rangle+|1111\rangle){}\nonumber + \frac{a-d}{2}(|1100\rangle+|0011\rangle){}\nonumber \\
& + & \frac{b+c}{2}(|0101\rangle+|1010\rangle)  {}\nonumber + \frac{b-c}{2}(|0110\rangle+|1001\rangle) 
\end{eqnarray}
with $a$, $b$, $c$ and $d$ being complex numbers and the normalization condition being $|a|^2+|b|^2+|c|^2+|d|^2 = 1$. Here, the first two qubits are with Alice, the third one with Bob and the last one with Charlie. Here Alice is the dealer who shares the state with both Bob and Charlie. The state to be shared by Alice is given by the one qubit state $|\psi_{S}\rangle = \alpha|0\rangle+\beta|1\rangle$ where $|\alpha|^2+|\beta|^2=1$. The combined state of the system with the message and the resource is given by:
\begin{eqnarray}
&&|\psi_{S}\rangle\otimes |G_{abcd}\rangle  = \{\alpha|0\rangle+\beta|1\rangle\}\otimes \frac{a+d}{2}(|0000\rangle+|1111\rangle) {}\nonumber \\ & + &\frac{a-d}{2}(|1100\rangle+|0011\rangle)  + \frac{b+c}{2}(|0101\rangle+|1010\rangle) {}\nonumber \\ & + &\frac{b-c}{2} (|0110\rangle+|1001\rangle) 
\end{eqnarray}
%
\noindent Our next target is to see how the dealer Alice is revoking back the unknown state which is already shared with Bob and Charlie.  Revocation is important when Alice decides to change the state or Alice guesses that recipients are no longer trustworthy, or there is an update of state in higher level application using state sharing as a subroutine.
Now if we rewrite the combined system with the  qubit $S$ and the first qubit of Alice in the Bell basis we have:
\begin{widetext}
\begin{eqnarray}
|\psi_{S}\rangle\otimes |G_{abcd}\rangle 
&=&(|\phi^+\rangle)[(\frac{a+d}{2\sqrt{2}})(\alpha|000\rangle+\beta|111\rangle)+(\frac{a-d}{2\sqrt{2}})(\alpha|011\rangle+\beta|100\rangle) + (\frac{b+c}{2\sqrt{2}})(\alpha|101\rangle+\beta|010\rangle){}\nonumber\\&&+(\frac{b-c}{2\sqrt{2}})(\alpha|110\rangle+\beta|001\rangle)] + (|\phi^-\rangle)[(\frac{a+d}{2\sqrt{2}})(\alpha|000\rangle-\beta|111\rangle)+(\frac{a-d}{2\sqrt{2}})(\alpha|011\rangle-\beta|100\rangle){}\nonumber\\&& + (\frac{b+c}{2\sqrt{2}})(\alpha|101\rangle-\beta|010\rangle)+(\frac{b-c}{2\sqrt{2}})(\alpha|110\rangle-\beta|001\rangle)]  {}\nonumber\\&&+(|\psi^+\rangle)[(\frac{a+d}{2\sqrt{2}})(\alpha|111\rangle+\beta|000\rangle)+(\frac{a-d}{2\sqrt{2}})(\alpha|100\rangle+\beta|011\rangle) + (\frac{b+c}{2\sqrt{2}})(\alpha|010\rangle+\beta|101\rangle){}\nonumber\\&&+(\frac{b-c}{2\sqrt{2}})(\alpha|001\rangle+\beta|110\rangle)] + (|\psi^-\rangle)[(\frac{a+d}{2\sqrt{2}})(\alpha|111\rangle-\beta|000\rangle)+(\frac{a-d}{2\sqrt{2}})(\alpha|100\rangle-\beta|011\rangle)  {}\nonumber\\&&+(\frac{b+c}{2\sqrt{2}})(\alpha|010\rangle-\beta|101\rangle)+(\frac{b-c}{2\sqrt{2}})(\alpha|001\rangle-\beta|110\rangle)]
\end{eqnarray}
\end{widetext}
{\center
\begin{table*}
\begin{center}
\begin{tabular}{|c|c|}
\hline \parbox{1.8in}{\begin{center}Alice's Measurement Outcomes\end{center}} & \parbox{1.8in}{\begin{center}Alice, Bob and Charlie's Joint State\end{center}}\\
\hline $\lvert\phi^\pm \rangle$  &
$\frac{a+d}{\sqrt{2}}[\alpha|000\rangle\pm\beta|111\rangle]$+  $\frac{a-d}{\sqrt{2}}[\alpha|011\rangle\pm\beta|100\rangle]$+  $\frac{b+c}{\sqrt{2}}[\alpha|101\rangle\pm\beta|010\rangle]$+  $\frac{b-c}{\sqrt{2}}[\alpha|110\rangle\pm\beta|001\rangle]$  \\
\hline $\lvert\psi^\pm\rangle$ & $\frac{a+d}{\sqrt{2}}[\alpha|111\rangle\pm\beta|000\rangle]$+ $\frac{a-d}{\sqrt{2}}[\alpha|100\rangle\pm\beta|011\rangle]$+  $\frac{b+c}{\sqrt{2}}[\alpha|010\rangle\pm\beta|101\rangle]$+ $\frac{b-c}{\sqrt{2}}[\alpha|001\rangle\pm\beta|110\rangle]$   \\
\hline
\end{tabular} 
\end{center}
\caption{Sharing of Quantum Information}
\label{tab:second_table}
\end{table*}
}
\begin{table*}
\begin{minipage}{.4\linewidth}
\label{tab:title} 
\begin{tabular}{|c|c|c|c|}
\hline \parbox{.6in}{\ \\Alice's  Outcomes\\ \ \\} & \parbox{0.6in}{\ \\Bob \& Charlie's Outcomes\\ \ \\} &\parbox{0.6in}{\ \\Alice's Resultant state\\ \ \\} & \parbox{0.6in}{\ \\Alice's Local Operations\ \\ \ \\}\\
\hline $|\phi^+\rangle$  & $|++\rangle$ & $\alpha|0\rangle+\beta|1\rangle$ & $I$    \\
\hline $|\phi^+\rangle$  & $|+-\rangle$ & $\alpha|0\rangle-\beta|1\rangle$ & $\sigma_z$  \\
\hline $|\phi^+\rangle$  & $|-+\rangle$ & $\alpha|0\rangle-\beta|1\rangle$ & $\sigma_z$  \\
\hline $|\phi^+\rangle$  & $|--\rangle$ & $\alpha|0\rangle+\beta|1\rangle$ & $I$  \\
\hline $|\phi^-\rangle$  & $|++\rangle$ & $\alpha|0\rangle-\beta|1\rangle$ & $\sigma_z$  \\
\hline $|\phi^-\rangle$  & $|+-\rangle$ & $\alpha|0\rangle+\beta|1\rangle$ & $I$ \\
\hline $|\phi^-\rangle$  & $|-+\rangle$ & $\alpha|0\rangle+\beta|1\rangle$ & $I$ \\
\hline $|\phi^-\rangle$  & $|--\rangle$ & $\alpha|0\rangle-\beta|1\rangle$ & $\sigma_z$ \\
\hline $|\psi^+\rangle$  & $|++\rangle$ & $\alpha|1\rangle+\beta|0\rangle$ & $\sigma_x$    \\
\hline $|\psi^+\rangle$  & $|+-\rangle$ & $-\alpha|1\rangle+\beta|0\rangle$ & $\sigma_x\sigma_z$  \\
\hline $|\psi^+\rangle$  & $|-+\rangle$ & $-\alpha|1\rangle+\beta|0\rangle$ & $\sigma_x\sigma_z$    \\
\hline $|\psi^+\rangle$  & $|--\rangle$ & $\alpha|1\rangle+\beta|0\rangle$ & $\sigma_x$    \\
\hline $|\psi^-\rangle$  & $|++\rangle$ & $\alpha|1\rangle-\beta|0\rangle$ & $\sigma_x\sigma_z$    \\
\hline $|\psi^-\rangle$  & $|+-\rangle$ & $-\alpha|1\rangle-\beta|0\rangle$ & $\sigma_x$  \\
\hline $|\psi^-\rangle$  & $|-+\rangle$ & $-\alpha|1\rangle-\beta|0\rangle$ & $\sigma_x$    \\
\hline $|\psi^-\rangle$  & $|--\rangle$ & $\alpha|1\rangle-\beta|0\rangle$ & $\sigma_x\sigma_z$    \\
\hline
\end{tabular} 
\end{minipage}\begin{minipage}{.8\linewidth}
\begin{tabular}{|c|c|c|c|}
\hline \parbox{.6in}{\ \\Alice's  Outcomes\\ \ \\} & \parbox{.6in}{\ \\Alice and Bob's Outcomes\\ \ \\} &\parbox{.6in}{\ \\Charlie's Resultant state\\ \ \\} & \parbox{.6in}{\ \\Charlie's Local Operations\ \\ \ \\}\\
\hline $|\phi^+\rangle$  & $|++\rangle$ & $\alpha|0\rangle+\beta|1\rangle$ & $I$    \\
\hline $|\phi^+\rangle$  & $|+-\rangle$ & $\alpha|0\rangle-\beta|1\rangle$ & $\sigma_z$  \\
\hline $|\phi^+\rangle$  & $|-+\rangle$ & $\alpha|0\rangle-\beta|1\rangle$ & $\sigma_z$  \\
\hline $|\phi^+\rangle$  & $|--\rangle$ & $\alpha|0\rangle+\beta|1\rangle$ & $I$  \\
\hline $|\phi^-\rangle$  & $|++\rangle$ & $\alpha|0\rangle-\beta|1\rangle$ & $\sigma_z$  \\
\hline $|\phi^-\rangle$  & $|+-\rangle$ & $\alpha|0\rangle+\beta|1\rangle$ & $I$ \\
\hline $|\phi^-\rangle$  & $|-+\rangle$ & $\alpha|0\rangle+\beta|1\rangle$ & $I$ \\
\hline $|\phi^-\rangle$  & $|--\rangle$ & $\alpha|0\rangle-\beta|1\rangle$ & $\sigma_z$ \\
\hline $|\psi^+\rangle$  & $|++\rangle$ & $\alpha|1\rangle+\beta|0\rangle$ & $\sigma_x$    \\
\hline $|\psi^+\rangle$  & $|+-\rangle$ & $-\alpha|1\rangle+\beta|0\rangle$ & $\sigma_x\sigma_z$  \\
\hline $|\psi^+\rangle$  & $|-+\rangle$ & $-\alpha|1\rangle+\beta|0\rangle$ & $\sigma_x\sigma_z$    \\
\hline $|\psi^+\rangle$  & $|--\rangle$ & $\alpha|1\rangle+\beta|0\rangle$ & $\sigma_x$    \\
\hline $|\psi^-\rangle$  & $|++\rangle$ & $\alpha|1\rangle-\beta|0\rangle$ & $\sigma_x\sigma_z$    \\
\hline $|\psi^-\rangle$  & $|+-\rangle$ & $-\alpha|1\rangle-\beta|0\rangle$ & $\sigma_x$  \\
\hline $|\psi^-\rangle$  & $|-+\rangle$ & $-\alpha|1\rangle-\beta|0\rangle$ & $\sigma_x$    \\
\hline $|\psi^-\rangle$  & $|--\rangle$ & $\alpha|1\rangle-\beta|0\rangle$ & $\sigma_x\sigma_z$    \\
\hline
\end{tabular} 
\end{minipage}
\caption{ 1. (LHS) Retrieving Quantum Secret from Alice's resultant state when ($b = 0$ and $c = 0$) and ($a =\frac{1}{\sqrt{2}}$ and $d = {\frac{1}{\sqrt{2}}}$). 2. (RHS) Reconstructing Quantum Secret from Charlie's resultant state when ($b = 0$ and $c = 0$) and ($a =\frac{1}{\sqrt{2}}$ and $d =\frac{1}{\sqrt{2}}$)}
\label{tab:third_table}
\end{table*}
Now Alice measures both the qubit to be shared and the first qubit of the resource state in the Bell basis. There will be four possible measurement outcomes $\{\phi^\pm, \psi^\pm\}$ as a result of Alice's measurement. In correspondence to various measurement outcomes obtained by Alice in the Bell basis, Alice, Bob and Charlie's qubits collapse into the states as given in \cref{tab:second_table}.
It is important to note that the information about the state is shared among three parties Alice, Bob and Charlie and Alice being the dealer also has a hold on the part of the state. The Alice's qubit that is shared in form of a three qubit state with Bob and Charlie is the \textbf{quantum share} of the state. Next we see how the dealer Alice is revoking it back which is already shared with Bob and Charlie.\\
%
%
\subsection{Revocation of State on Alice's location} 
\noindent Now if we write Bob and Charlie's qubit in the Hadamard basis, then the combined shared system of Alice, Bob and Charlie (with Bob's and Charlie's qubit are written in $|+\rangle$ and $|-\rangle$ basis), when Alice's initial measurement outcome is $|\phi^{\pm} \rangle$, can be rewritten as:
\begin{widetext}
\begin{eqnarray}
|Y_{234}^{\phi^\pm}\rangle&=&
(\frac{a+d}{2\sqrt{2}})\{\alpha[|0\rangle_{A}(|+\rangle+|-\rangle)_{B}{} (|+\rangle+|-\rangle)_{C}]\pm{}\beta[|1\rangle_{A}(|+\rangle-|-\rangle)_{B}(|+\rangle-|-\rangle)_{C}]\}{}\nonumber\\&+&(\frac{a-d}{2\sqrt{2}})\{\alpha[|0\rangle_{A}(|+\rangle+|-\rangle)_{B} (|+\rangle+|-\rangle)_{C}]\pm{}\beta[|1\rangle_{A}(|+\rangle-|-\rangle)_{B} (|+\rangle-|-\rangle)_{C}]\}{}\nonumber\\&+&(\frac{b+c}{2\sqrt{2}})\{[\alpha[|1\rangle_{A}(|+\rangle+|-\rangle)_{B} (|+\rangle-|-\rangle)_{C}]|1\rangle_{C}{}\pm{}\beta[|0\rangle_{A}(|+\rangle+|-\rangle)_{B} (|+\rangle+|-\rangle)_{C}]\} {}\nonumber\\&+&(\frac{b-c}{2\sqrt{2}})\{[\alpha[|1\rangle_{A}(|+\rangle-|-\rangle)_{A} (|+\rangle+|-\rangle)_{B}|0\rangle_{C}]\pm{}\beta[|0\rangle_{A}(|+\rangle+|-\rangle)_{B} (|+\rangle-|-\rangle)_{C}]\}
\end{eqnarray}
\noindent And if we write Bob and Charlie's qubit in the Hadamard basis, then the combined shared system of Alice, Bob and Charlie, when Alice's initial measurement outcome is $|\psi^{\pm} \rangle$ can be rewritten as,
\begin{eqnarray}
|Y_{234}^{\psi^\pm}\rangle&=&
(\frac{a+d}{2\sqrt{2}})\{\alpha[|1\rangle_{A}(|+\rangle-|-\rangle)_{B}(|+\rangle-|-\rangle)_{C}]\pm{}\beta[|0\rangle_{A}(|+\rangle+|-\rangle)_{B}(|+\rangle+|-\rangle)_{C}]\}{}\nonumber\\&+&(\frac{a-d}{2\sqrt{2}})\{\alpha[|1\rangle_{A}(|+\rangle+|-\rangle)_{B} (|+\rangle+|-\rangle)_{C}]\pm{}\beta[|0\rangle_{A}(|+\rangle-|-\rangle)_{B} (|+\rangle-|-\rangle)_{C}\}{}\nonumber\\&+&(\frac{b+c}{2\sqrt{2}})\{[\alpha[|0\rangle_{A}(|+\rangle-|-\rangle)_{B} (|+\rangle+|-\rangle)_{C}]|1\rangle_{C}\pm{}\beta[|1\rangle_{A}(|+\rangle+|-\rangle)_{B} (|+\rangle-|-\rangle)_{C}]\} {}\nonumber\\&+&(\frac{b-c}{2\sqrt{2}})\{[\alpha[|0\rangle_{A}(|+\rangle+|-\rangle)_{A} (|+\rangle-|-\rangle)_{B}]\pm{}\beta[|1\rangle_{A}(|+\rangle-|-\rangle)_{B} (|+\rangle0|-\rangle)_{C}]\}
\end{eqnarray}
\end{widetext}
%
\noindent Alice’s resultant state after Bob and Charlie measure their respective qubits in Hadamard Basis (when Alice's initial measurement outcome is $|\phi^{\pm} \rangle$):\\

\noindent \textbf{Case 1:} When Bob measures his qubit in Hadamard Basis $|+\rangle$ and Charlie measures his respective qubit in the Hadamard Basis $|+\rangle$, Alice's resultant state is as follows:
$\sqrt{2}(a\alpha\pm b\beta)|0\rangle+\sqrt{2}(b\alpha\pm a\beta)|1\rangle.$\\

\noindent \textbf{Case 2:}  When Bob measures his qubit in Hadamard Basis $|+\rangle$ and Charlie measures his respective qubit in the Hadamard Basis $|-\rangle$, Alice's resultant state is as follows:
$\sqrt{2}(d\alpha \pm c\beta)|0\rangle-\sqrt{2}(c\alpha \pm d\beta)|1\rangle.$\\ 

\noindent \textbf{Case 3:}  When Bob measures his qubit in Hadamard Basis $|-\rangle$ and Charlie measures his respective qubit in the Hadamard Basis $|+\rangle$, Alice's resultant state is as follows:
$\sqrt{2}(d\alpha \mp c\beta)|0\rangle+\sqrt{2}(c\alpha \mp d\beta)|1\rangle. $\\

\noindent \textbf{Case 4:}  When Bob measures his qubit in Hadamard Basis $|-\rangle$ and Charlie measures his respective qubit in the Hadamard Basis $|-\rangle$, Alice's resultant state is as follows:
$\sqrt{2}(a\alpha \mp b\beta)|0\rangle-\sqrt{2}(b\alpha \mp a\beta)|1\rangle. $\\

\noindent Alice’s resultant state after Bob and Charlie measure their respective qubits in Hadamard Basis (when Alice's initial measurement outcome is $|\psi^{\pm} \rangle$):

 \noindent \textbf{Case 1:} When Bob measures his qubit in Hadamard Basis $|+\rangle$ and Charlie measures his respective qubit in the Hadamard Basis $|-\rangle$, Alice's resultant state is as follows:
$\sqrt{2}(b\alpha\pm a\beta)|0\rangle+\sqrt{2}(a\alpha\pm b\beta)|1\rangle.$\\

\noindent \textbf{Case 2:} When Bob measures his qubit in Hadamard Basis $|+\rangle$ and Charlie measures his respective qubit in the Hadamard Basis $|-\rangle$, Alice's resultant state is as follows:
$\sqrt{2}(c\alpha \pm d\beta)|0\rangle-\sqrt{2}(d\alpha \pm c\beta)|1\rangle.$\\ 

\noindent \textbf{Case 3:} When Bob measures his qubit in Hadamard Basis $|-\rangle$ and Charlie measures his respective qubit in the Hadamard Basis $|+\rangle$, Alice's resultant state is as follows:
$-\sqrt{2}(c\alpha \mp d\beta)|0\rangle-\sqrt{2}(d\alpha \mp c\beta)|1\rangle. $\\

\noindent \textbf{Case 4:} When Bob measures his qubit in Hadamard Basis $|-\rangle$ and Charlie measures his respective qubit in the Hadamard Basis $|-\rangle$, Alice's resultant state is as follows:
$-\sqrt{2}(b\alpha \mp a\beta)|0\rangle-\sqrt{2}(a\alpha \mp b\beta)|1\rangle $.\\

\noindent Now if Bob and Charlie send their measurement outcomes in the form of two cbits with the encoding  $|+\rangle \rightarrow |0\rangle$, $|-\rangle \rightarrow |1\rangle$, Alice will be able to revive the state with the applications of appropriate Pauli operator. However this will not be possible for all values of $a,b,c,d$. There will be certain range of the input parameters for which revocation will be possible. Next in the form of following theorem, we give the range of input parameters for which Alice will be able to revive.   
\begin{theorem}
For  ${G}_{abcd}$ state, the necessary conditions for Alice to retrieve the state are as follows:\\

     \noindent 1. If $a$ and $c$ are chosen as real numbers, then we have $b$ and $d$ as purely imaginary numbers such that: \\
     (a) $a^2 \leq \frac{1}{2}$ or $-\frac{1}{\sqrt{2}} \leq a \leq \frac{1}{\sqrt{2}}$ (b)\ $c^2 \leq \frac{1}{2}$ or $-\frac{1}{\sqrt{2}} \leq c \leq \frac{1}{\sqrt{2}}$
     (c)\ $|b|^2 = \frac{1}{2} - a^2$ 
     (d)\ $|d|^2 = \frac{1}{2} - c^2$\\
     
     \noindent 2. If $a$ and $d$ are chosen to be real numbers, then we have $b$ and $c$ as purely imaginary numbers such that:
     (a) $a^2 \leq \frac{1}{2}$ or  $-\frac{1}{\sqrt{2}} \leq a \leq \frac{1}{\sqrt{2}}$
     (b) $d^2 \leq \frac{1}{2}$ or $-\frac{1}{\sqrt{2}} \leq d \leq \frac{1}{\sqrt{2}}$
     (c)\ $|b|^2 = \frac{1}{2} - a^2$ 
     (d)\ $|c|^2 = \frac{1}{2} - d^2$\\ 
     
     \noindent 3. If $b$ and $c$ are chosen to be real numbers, then we have $a$ and $d$ as purely imaginary numbers such that:
     (a) $b^2 \leq \frac{1}{2}$ or $-\frac{1}{\sqrt{2}} \leq b \leq \frac{1}{\sqrt{2}}$
     (b)\ $c^2 \leq \frac{1}{2}$ or $-\frac{1}{\sqrt{2}} \leq c \leq \frac{1}{\sqrt{2}}$
     (c) $|a|^2 = \frac{1}{2} - b^2$ 
     (d)\ $|d|^2 = \frac{1}{2} - c^2$\\
     
     \noindent 4. If $b$ and $d$ are chosen to be real numbers, then we have $a$ and $c$ as purely imaginary numbers such that:
     (a) $b^2 \leq \frac{1}{2}$ or $-\frac{1}{\sqrt{2}} \leq b \leq \frac{1}{\sqrt{2}}$
     (b) $d^2 \leq \frac{1}{2}$ or $-\frac{1}{\sqrt{2}} \leq d \leq \frac{1}{\sqrt{2}}$
     (c) $|a|^2 = \frac{1}{2} - b^2$ 
     (d) $|c|^2 = \frac{1}{2} - d^2$\\
     
\end{theorem}
\noindent The union of conditions 1, 2, 3 and 4 stated in Theorem 1 given above gives all the necessary values of a, b, c and d for Alice to retrieve the state.
For the protocol to be successful, the reconstruction at the shareholder's place should be possible in the normal scenario when all the share holder's are not dishonest. Next we consider the reconstruction of the state at the Charlie's place (say). The same procedure can be repeated for Bob also. 
\subsection{Reconstruction of State on Charlie's location} Now if we write Alice and Bob's qubit in the Hadamard basis, then the combined shared system of Alice, Bob and Charlie, when Alice's initial measurement is $|\phi^{\pm}\rangle$ can be rewritten as:  
\begin{eqnarray}
|Y_{234}^{\phi^{\pm}}\rangle&=&
(\frac{a+d}{2\sqrt{2}})\{\alpha[(|+\rangle+|-\rangle)_{A} (|+\rangle+|-\rangle)_{B}|0\rangle_{C}]{}\nonumber\\&&\pm\beta[(|+\rangle-|-\rangle)_{A}(|+\rangle-|-\rangle)_{B}|1\rangle_{C}]\}{}\nonumber\\&&+(\frac{a-d}{2\sqrt{2}})\{\alpha[(|+\rangle+|-\rangle)_{A} (|+\rangle-|-\rangle)_{B}|1\rangle_{C}]{}\nonumber\\&&\pm\beta[(|+\rangle-|-\rangle)_{A} (|+\rangle+|-\rangle)_{B}|0\rangle_{C}]\}{}\nonumber\\&&+(\frac{b+c}{2\sqrt{2}})\{[\alpha[(|+\rangle-|-\rangle)_{A} (|+\rangle+|-\rangle)_{B}]|1\rangle_{C}{}\nonumber\\&&\pm\beta[(|+\rangle+|-\rangle)_{A} (|+\rangle-|-\rangle)_{B}|0\rangle_{C}] {}\nonumber\\&&+(\frac{b-c}{2\sqrt{2}})\{[\alpha[(|+\rangle-|-\rangle)_{A} (|+\rangle-|-\rangle)_{B}|0\rangle_{C}]{}\nonumber\\&&\pm\beta[(|+\rangle+|-\rangle)_{A} (|+\rangle+|-\rangle)_{B}|1\rangle_{C}]\}
\end{eqnarray}
The combined shared system, when Alice's initial measurement is $|\psi^{\pm}\rangle$ is given by 
\begin{eqnarray}
|Y_{234}^{\psi^{\pm}}\rangle&=&
(\frac{a+d}{2\sqrt{2}})\{\alpha[(|+\rangle-|-\rangle)_{A} (|+\rangle-|-\rangle)_{B}|1\rangle_{C}]\pm{}\nonumber\\&&\beta[(|+\rangle+|-\rangle)_{A} (|+\rangle+|-\rangle)_{B}|0\rangle_{C}]\}+{}\nonumber\\&&(\frac{a-d}{2\sqrt{2}})\{\alpha[(|+\rangle-|-\rangle)_{A} (|+\rangle+|-\rangle)_{B}|0\rangle_{C}]\pm{}\nonumber\\&&\beta[(|+\rangle+|-\rangle)_{A} (|+\rangle-|-\rangle)_{B}|1\rangle_{C}]\}+{}\nonumber\\&&(\frac{b+c}{2\sqrt{2}})\{[\alpha[(|+\rangle+|-\rangle)_{A} (|+\rangle-|-\rangle)_{B}]|0\rangle_{C}\pm{}\nonumber\\&&\beta[(|+\rangle-|-\rangle)_{A} (|+\rangle+|-\rangle)_{B}|1\rangle_{C}] -{}\nonumber\\&&(\frac{b-c}{2\sqrt{2}})\{[\alpha[(|+\rangle+|-\rangle)_{A} (|+\rangle+|-\rangle)_{B}|1\rangle_{C}]\pm{}\nonumber\\&&\beta[(|+\rangle-|-\rangle)_{A} (|+\rangle-|-\rangle)_{B}|0\rangle_{C}]\}. 
\end{eqnarray}

\noindent Charlie's resultant state after Alice and Bob measure their respective qubits in Hadamard Basis (when Alice's initial measurement outcome is $\lvert\phi^{\pm}\rangle$) :\\

\noindent \textbf{Case 1:} When Alice measures her qubit in Hadamard Basis $|+\rangle$ and Bob measures his respective qubit in the Hadamard Basis $|+\rangle$, Charlie's resultant state is as follows: $\frac{1}{\sqrt{2}}[[(a+b-c+d)\alpha \pm (a+b+c-d)\beta)]|0\rangle+[(a+b+c-d)\alpha \pm (a+b-c+d)\beta)]|1\rangle]$\\ 

\noindent \textbf{Case 2:} When Alice measures her qubit in Hadamard Basis $|+\rangle$ and Bob measures his respective qubit in the Hadamard Basis $|-\rangle$, Charlie's resultant state is as follows: $\frac{1}{\sqrt{2}}[[(a-b+c+d)\alpha \pm (a-b-c-d)\beta)]|0\rangle+[(-a+b+c+d)\alpha \pm (-a+b-c-d)\beta)]|1\rangle]$\\

\noindent \textbf{Case 3:} When Alice measures her qubit in Hadamard Basis $|-\rangle$ and Bob measures his respective qubit in the Hadamard Basis $|+\rangle$, Charlie's resultant state is as follows:
$\frac{1}{\sqrt{2}}[[(a-b+c+d)\alpha \pm  (-a+b+c+d)\beta)]|0\rangle+[(a-b-c-d)\alpha \pm (-a+b-c-d)\beta)]|1\rangle] $\\

\noindent \textbf{Case 4:} When Alice measures her qubit in Hadamard Basis $|-\rangle$ and Bob measures his respective qubit in the Hadamard Basis $|-\rangle$, Charlie's resultant state is as follows:
$\frac{1}{\sqrt{2}}[[(a+b-c+d)\alpha \pm (-a-b-c+d)\beta)]|0\rangle+[(-a-b-c+d)\alpha \pm (a+b-c+d)\beta)]|1\rangle] $\\

\noindent Charlie's resultant state after Alice and Bob measure their respective qubits in Hadamard Basis (when Alice's initial measurement outcome is $|\psi^{\pm}\rangle$) :\\

\noindent \textbf{Case 1:}  When Alice measures her qubit in Hadamard Basis $|+\rangle$ and Bob measures his respective qubit in the Hadamard Basis $|+\rangle$, Charlie's resultant state is as follows: $\frac{1}{\sqrt{2}}[[(a+b+c-d)\alpha \pm (a+b-c+d)\beta)]|0\rangle+[(a+b-c+d)\alpha \pm (a+b+c-d)\beta)]|1\rangle].$\\ 

\noindent \textbf{Case 2:} When Alice measures her qubit in Hadamard Basis $|+\rangle$ and Bob measures his respective qubit in the Hadamard Basis $|-\rangle$, Charlie's resultant state is as follows: $\frac{1}{\sqrt{2}}[[(a-b-c-d)\alpha \pm (a-b+c+d)\beta)]|0\rangle+[(-a+b-c-d)\alpha \pm (-a+b+c+d)\beta)]|1\rangle].$\\

\noindent \textbf{Case 3:} When Alice measures her qubit in Hadamard Basis $|-\rangle$ and Bob measures his respective qubit in the Hadamard Basis $|+\rangle$, Charlie's resultant state is as follows:
$\frac{1}{\sqrt{2}}[[(-a+b+c+d)\alpha \pm (a-b+c+d)\beta)]|0\rangle+[(-a+b-c-d)\alpha \pm (a-b-c-d)\beta)]|1\rangle]. $\\

\noindent \textbf{Case 4:} When Alice measures her qubit in Hadamard Basis $|-\rangle$ and Bob measures his respective qubit in the Hadamard Basis $|-\rangle$, Charlie's resultant state is as follows:
$\frac{1}{\sqrt{2}}[[(-a-b-c+d)\alpha \pm (a+b-c+d)\beta)]|0\rangle+[(a+b-c+d)\alpha \pm (-a-b-c+d)\beta)]|1\rangle]. $\\



\noindent Now if Alice and Bob send their measurement outcomes in the form of two cbits with the same encoding $|+\rangle \rightarrow |0\rangle$, $|-\rangle \rightarrow |1\rangle$, Charlie can reconstruct the initial state by applying appropriate Pauli operators on his obtained state. Just like in the case of revocation, here reconstruction of the state will not be possible for all values of $a,b,c,d$. There will be certain range of the input parameters for which the reconstruction will take place. Next in the following theorem, we give the range of input parameters for which Charlie will be able to reconstruct the state. \\

\begin{theorem}
For the protocol to be implemented using the ${G}_{abcd}$ state, the necessary conditions for Charlie to reconstruct the state are as follows:\\

     \noindent 
     1. If $a$ and $c$ are chosen as real numbers, and $b$ and $d$ are chosen as imaginary numbers, then we have:\\
     (a) {$a = c$ and $b = d$}
     (b) $a^2 \leq \frac{1}{2}$ or $-\frac{1}{\sqrt{2}} \leq a \leq \frac{1}{\sqrt{2}}$
     (c) $c^2 \leq \frac{1}{2}$ or $-\frac{1}{\sqrt{2}} \leq c \leq \frac{1}{\sqrt{2}}$
     (d) $|b|^2 = \frac{1}{2} - a^2$ 
     (e) $|d|^2 = \frac{1}{2} - c^2$ 
     \noindent 2. If $a$ and $d$ are chosen as real numbers, and $b$ and $c$ are chosen as imaginary numbers, then we have:\\ 
     (a) {$a = -d$ and $b = -c$}
     (b) $a^2 \leq \frac{1}{2}$ or $-\frac{1}{\sqrt{2}} \leq a \leq \frac{1}{\sqrt{2}}$
     (c) $d^2 \leq \frac{1}{2}$ or $-\frac{1}{\sqrt{2}} \leq d \leq \frac{1}{\sqrt{2}}$
     (d) $|b|^2 = \frac{1}{2} - a^2$ 
     (e) $|c|^2 = \frac{1}{2} - d^2$\\
     
     \noindent 
     3. If $b$ and $c$ are chosen as real numbers, and $a$ and $d$ are chosen as imaginary numbers, then we have:\\ 
     (a) {$b = c$ and $a = d$}
     (b) $b^2 \leq \frac{1}{2}$ or $-\frac{1}{\sqrt{2}} \leq b \leq \frac{1}{\sqrt{2}}$
     (c) $c^2 \leq \frac{1}{2}$ or $-\frac{1}{\sqrt{2}} \leq c \leq \frac{1}{\sqrt{2}}$
     (d) $|a|^2 = \frac{1}{2} - b^2$ 
     (e) $|d|^2 = \frac{1}{2} - c^2$ 
     \noindent 
     4. If $b$ and $d$ are chosen as real numbers, and $a$ and $c$ are chosen as imaginary numbers, then we have:\\ 
     (a) {$b = -d$ and $a = -c$}
     (b) $b^2 \leq \frac{1}{2}$ or $-\frac{1}{\sqrt{2}} \leq b \leq \frac{1}{\sqrt{2}}$
     (c) $d^2 \leq \frac{1}{2}$ or $-\frac{1}{\sqrt{2}} \leq d \leq \frac{1}{\sqrt{2}}$
     (d) $|a|^2 = \frac{1}{2} - b^2$ 
     (e) $|c|^2 = \frac{1}{2} - d^2$\\ 
     
     \noindent 
     5. If $a$ and $c$ are chosen as real numbers, and $b$ and $d$ are chosen as imaginary numbers, then we have:\\
     (a) {$b = -d$ and $a = -c$}
     (b) $a^2 \leq \frac{1}{2}$ or $-\frac{1}{\sqrt{2}} \leq a \leq \frac{1}{\sqrt{2}}$
     (c) $c^2 \leq \frac{1}{2}$ or $-\frac{1}{\sqrt{2}} \leq c \leq \frac{1}{\sqrt{2}}$
     (d) $|b|^2 = \frac{1}{2} - a^2$ 
     (e) $|d|^2 = \frac{1}{2} - c^2$ 
     \noindent 6. If $a$ and $d$ are chosen as real numbers, and $b$ and $c$ are chosen as imaginary numbers, then we have:\\ 
     (a) {$b = c$ and $a = d$}
     (b) $a^2 \leq \frac{1}{2}$ or $-\frac{1}{\sqrt{2}} \leq a \leq \frac{1}{\sqrt{2}}$
     (c) $d^2 \leq \frac{1}{2}$ or $-\frac{1}{\sqrt{2}} \leq d \leq \frac{1}{\sqrt{2}}$
     (d) $|b|^2 = \frac{1}{2} - a^2$ 
     (e) $|c|^2 = \frac{1}{2} - d^2$\\
     
     \noindent 
     7. If $b$ and $c$ are chosen as real numbers, and $a$ and $d$ are chosen as imaginary numbers, then we have:\\ 
     (a) {$a = -d$ and $b = -c$}
     (b) $b^2 \leq \frac{1}{2}$ or $-\frac{1}{\sqrt{2}} \leq b \leq \frac{1}{\sqrt{2}}$
     (c) $c^2 \leq \frac{1}{2}$ or $-\frac{1}{\sqrt{2}} \leq c \leq \frac{1}{\sqrt{2}}$
     (d) $|a|^2 = \frac{1}{2} - b^2$ 
     (e) $|d|^2 = \frac{1}{2} - c^2$\\
     
     \noindent 
     8. If $b$ and $d$ are chosen as real numbers, and $a$ and $c$ are chosen as imaginary numbers, then we have:\\ 
     (a) {$a = c$ and $b = d$}
     (b) $b^2 \leq \frac{1}{2}$ or $-\frac{1}{\sqrt{2}} \leq b \leq \frac{1}{\sqrt{2}}$
     (c) $d^2 \leq \frac{1}{2}$ or $-\frac{1}{\sqrt{2}} \leq d \leq \frac{1}{\sqrt{2}}$
     (d) $|a|^2 = \frac{1}{2} - b^2$ 
     (e) $|c|^2 = \frac{1}{2} - d^2$\\

\end{theorem}
\noindent For the proposed protocol to be successful there will be common range of the values of $a,b,c,d$ for which both revocation and reconstruction will be possible which we get by the union of conditions 1, 2, 3, 4, 5, 6, 7 and 8 in theorem 2 given above. 
\\ \\
\noindent \textit{\bf Note:} For proof of \textbf{Theorem 1} and \textbf{Theorem 2}, please see the Appendix.
\\
\section{Conclusion:} In this article we  address the revocation and reconstruction of quantum state for  1-dealer and 2-share holders scenario  using 4-qubit entangled states. Here Alice, (the dealer) also possesses a \textbf{quantum share}  of the  state, and can always bring back her state with the help of that quantum share, when she finds both shareholders to be dishonest, but with the help of Bob and Charlie. Revocation is important when Alice decides
to change the state or Alice guesses that recipients are no longer trustworthy, or there is an update of state in higher level application using state sharing as a subroutine. The protocol ensures the normal reconstruction of the state at shareholder's location otherwise. The actual advantage of the protocol is to prevent the  shareholders from any possibility of
reconstructing the state, with a probability of $1/4$, if they are  dishonest. We not only give the basic idea of the protocol but also show the workings of  the protocol with the help of a class of four qubit state. In particular, we find out the explicit ranges based on the input state parameter for which our protocol will be successful. In addition to that within that range we take a state as an example to demonstrate our protocol. 

\bibliographystyle{unsrt}
\bibliography{ref.bib}

@article{zhang2020three,
  title={Three-party quantum private computation of cardinalities of set intersection and union based on GHZ states},
  author={Zhang, Cai and Long, Yinxiang and Sun, Zhiwei and Li, Qin and Huang, Qiong},
  journal={Scientific Reports},
  volume={10},
  number={1},
  pages={22246},
  year={2020},
  publisher={Nature Publishing Group UK London}
}

@article{sazim2015retrieving,
  title={Retrieving and routing quantum information in a quantum network},
  author={Sazim, Sk and Chiranjeevi, V and Chakrabarty, Indranil and Srinathan, Kannan},
  journal={Quantum Information Processing},
  volume={14},
  number={12},
  pages={4651--4664},
  year={2015},
  publisher={Springer}
}

@article{hillery1999quantum,
  title={Quantum secret sharing},
  author={Hillery, Mark and Bu{\v{z}}ek, Vladim{\'\i}r and Berthiaume, Andr{\'e}},
  journal={Physical Review A},
  volume={59},
  number={3},
  pages={1829},
  year={1999},
  publisher={APS}
}

@article{cleve1999share,
  title={How to share a quantum secret},
  author={Cleve, Richard and Gottesman, Daniel and Lo, Hoi-Kwong},
  journal={Physical review letters},
  volume={83},
  number={3},
  pages={648},
  year={1999},
  publisher={APS}
}

@article{bennett1993teleporting,
  title={Teleporting an unknown quantum state via dual classical and Einstein-Podolsky-Rosen channels},
  author={Bennett, Charles H and Brassard, Gilles and Cr{\'e}peau, Claude and Jozsa, Richard and Peres, Asher and Wootters, William K},
  journal={Physical review letters},
  volume={70},
  number={13},
  pages={1895},
  year={1993},
  publisher={APS}
}

@article{pati2023teleportation,
  title={Teleportation of quantum coherence},
  author={Pati, Arun K and Aradhya, Vijeth and Chakrabarty, Indranil and Patro, Subhasree and others},
  journal={Physical Review A},
  volume={108},
  number={4},
  pages={042620},
  year={2023},
  publisher={APS}
}

@article{shor2000simple,
  title={Simple proof of security of the BB84 quantum key distribution protocol},
  author={Shor, Peter W and Preskill, John},
  journal={Physical review letters},
  volume={85},
  number={2},
  pages={441},
  year={2000},
  publisher={APS}
}

@article{grilo2025security,
  title={Security of a secret sharing protocol on the Qline},
  author={Grilo, Alex B and Hanouz, Lucas and Marin, Anne},
  journal={arXiv preprint arXiv:2504.19702},
  year={2025}
}

@article{basak2025resource,
  title={Resource Reduction in Multiparty Quantum Secret Sharing of both Classical and Quantum Information under Noisy Scenario},
  author={Basak, Nirupam and Paul, Goutam},
  journal={arXiv preprint arXiv:2504.16709},
  year={2025}
}

@article{ma2025quantum,
  title={Quantum ($ t $, $ n $) Threshold Multi-Secret Sharing based on Cluster States},
  author={Ma, Rui-Hai and Chen, Hui-Nan and Cai, Bin-Bin and Lin, Song and Zhang, Xiao-Chen},
  journal={arXiv preprint arXiv:2505.09317},
  year={2025}
}

@article{zhang2025high,
  title={High-efficient long-distance device-independent quantum secret sharing based on single-photon sources},
  author={Zhang, Qi and Zhang, Cheng and Zhong, Wei and Du, Ming-Ming and Zhou, Lan and Sheng, Yu-Bo},
  journal={arXiv preprint arXiv:2505.10797},
  year={2025}
}

@article{chiwaki2025measurement,
  title={Measurement-free reconstruction circuit of quantum secrets in quantum secret sharing},
  author={Chiwaki, Shogo and Matsumoto, Ryutaroh},
  journal={arXiv preprint arXiv:2505.18840},
  year={2025}
}

@article{yan2025quantum,
  title={Quantum secret sharing in a triangular superconducting quantum network},
  author={Yan, Haoxiong and Zang, Allen and Grebel, Joel and Wu, Xuntao and Chou, Ming-Han and Andersson, Gustav and Conner, Christopher R and Joshi, Yash J and Li, Shiheng and Miller, Jacob M and others},
  journal={arXiv preprint arXiv:2506.10878},
  year={2025}
}

@article{ekert1991quantum,
  title={Quantum cryptography based on Bell’s theorem},
  author={Ekert, Artur K},
  journal={Physical review letters},
  volume={67},
  number={6},
  pages={661},
  year={1991},
  publisher={APS}
}

@article{pati2000minimum,
  title = {Minimum classical bit for remote preparation and measurement of a qubit},
	volume = {63},
	copyright = {http://link.aps.org/licenses/aps-default-license},
	issn = {1050-2947, 1094-1622},
	url = {https://link.aps.org/doi/10.1103/PhysRevA.63.014302},
	doi = {10.1103/PhysRevA.63.014302},
	number = {1},
	urldate = {2024-09-27},
	journal = {Physical Review A},
	author = {Pati, Arun K.},
	month = dec,
	year = {2000},
	pages = {014302},
}

@article{srivastava2019one,
  title={One-shot conclusive multiport quantum dense coding capacities},
  author={Srivastava, Chirag and Bera, Anindita and Sen, Aditi and Sen, Ujjwal},
  journal={Physical Review A},
  volume={100},
  number={5},
  pages={052304},
  year={2019},
  publisher={APS}
}

@article{sazim2013study,
  title={A study of teleportation and super dense coding capacity in remote entanglement distribution},
  author={Sazim, Sk and Chakrabarty, Indranil},
  journal={The European Physical Journal D},
  volume={67},
  pages={1--8},
  year={2013},
  publisher={Springer}
}

@article{sohail2023teleportation,
  title={Teleportation of quantum coherence},
  author={Sohail and Pati, Arun K and Aradhya, Vijeth and Chakrabarty, Indranil and Patro, Subhasree},
  journal={Physical Review A},
  volume={108},
  number={4},
  pages={042620},
  year={2023},
  publisher={APS}
}

@article{chakrabarty2010teleportation,
  title={Teleportation via a mixture of a two qubit subsystem of a N-qubit W and GHZ state},
  author={Chakrabarty, Indranil},
  journal={The European Physical Journal D},
  volume={57},
  number={2},
  pages={265--269},
  year={2010},
  publisher={Springer}
}

@article{bennett1992communication,
  title={Communication via one-and two-particle operators on Einstein-Podolsky-Rosen states},
  author={Bennett, Charles H and Wiesner, Stephen J},
  journal={Physical review letters},
  volume={69},
  number={20},
  pages={2881},
  year={1992},
  publisher={APS}
}

@article{singh2024controlled,
  title={Controlled state reconstruction and quantum secret sharing},
  author={Singh, Pahulpreet and Chakrabarty, Indranil},
  journal={Physical Review A},
  volume={109},
  number={3},
  pages={032406},
  year={2024},
  publisher={APS}
}

@article{adhikari2010probabilistic,
  title={Probabilistic secret sharing through noisy quantum channels},
  author={Adhikari, Satyabrata and Chakrabarty, Indranil and Agrawal, Pankaj},
  journal={arXiv preprint arXiv:1012.5570},
  year={2010}
}

@article{bell1964einstein,
  title={On the einstein podolsky rosen paradox},
  author={Bell, John S},
  journal={Physics Physique Fizika},
  volume={1},
  number={3},
  pages={195},
  year={1964},
  publisher={APS}
}

@article{abrol2024secret,
  title={Secret Sharing, Teleportation and Bell's Inequality},
  author={Abrol, Pratishtha and Singh, Pahulpreet and Chakrabarty, Indranil},
  journal={arXiv preprint arXiv:2404.01212},
  year={2024}
}

@article{garg2024estimation,
  title={Estimation of power in the controlled quantum teleportation through the witness operator},
  author={Garg, Anuma and Adhikari, Satyabrata},
  journal={The European Physical Journal D},
  volume={78},
  number={5},
  pages={64},
  year={2024},
  publisher={Springer}
}

@article{bennett2005remote,
  title={Remote preparation of quantum states},
  author={Bennett, Charles H and Hayden, Patrick and Leung, Debbie W and Shor, Peter W and Winter, Andreas},
  journal={IEEE Transactions on Information Theory},
  volume={51},
  number={1},
  pages={56--74},
  year={2005},
  publisher={IEEE}
}

@article{wehner2018quantum,
  title={Quantum internet: A vision for the road ahead},
  author={Wehner, Stephanie and Elkouss, David and Hanson, Ronald},
  journal={Science},
  volume={362},
  number={6412},
  pages={eaam9288},
  year={2018},
  publisher={American Association for the Advancement of Science}
}

@article{wootters1982single,
  title={A single quantum cannot be cloned},
  author={Wootters, William K and Zurek, Wojciech H},
  journal={Nature},
  volume={299},
  number={5886},
  pages={802--803},
  year={1982},
  publisher={Nature Publishing Group UK London}
}

@article{li2009generation,
  title={Generation and propagation of entanglement in driven coupled-qubit systems},
  author={Li, Jian and Paraoanu, GS},
  journal={New Journal of Physics},
  volume={11},
  number={11},
  pages={113020},
  year={2009},
  publisher={IOP Publishing}
}

@article{karlsson1999quantum,
  title={Quantum entanglement for secret sharing and secret splitting},
  author={Karlsson, Anders and Koashi, Masato and Imoto, Nobuyuki},
  journal={Physical Review A},
  volume={59},
  number={1},
  pages={162},
  year={1999},
  publisher={APS}
}

@article{bandyopadhyay2000teleportation,
  title={Teleportation and secret sharing with pure entangled states},
  author={Bandyopadhyay, Somshubhro},
  journal={Physical Review A},
  volume={62},
  number={1},
  pages={012308},
  year={2000},
  publisher={APS}
}

@article{markham2008graph,
  title={Graph states for quantum secret sharing},
  author={Markham, Damian and Sanders, Barry C},
  journal={Physical Review A},
  volume={78},
  number={4},
  pages={042309},
  year={2008},
  publisher={APS}
}

@article{li2010semiquantum,
  title={Semiquantum secret sharing using entangled states},
  author={Li, Qin and Chan, Wai Hong and Long, Dong-Yang},
  journal={Physical Review A},
  volume={82},
  number={2},
  pages={022303},
  year={2010},
  publisher={APS}
}

@article{bogdanski2008experimental,
  title={Experimental quantum secret sharing using telecommunication fiber},
  author={Bogdanski, Jan and Rafiei, Nima and Bourennane, Mohamed},
  journal={Physical Review A},
  volume={78},
  number={6},
  pages={062307},
  year={2008},
  publisher={APS}
}

@article{schmid2006experimental,
  title={Experimental quantum secret sharing},
  author={Schmid, Ch and Trojek, Pavel and Gaertner, Sascha and Bourennane, Mohamed and Kurtsiefer, Ch and Zukowski, Marek and Weinfurter, Harald},
  journal={Fortschritte der Physik: Progress of Physics},
  volume={54},
  number={8-10},
  pages={831--839},
  year={2006},
  publisher={Wiley Online Library}
}

@article{schmid2005experimental,
  title={Experimental single qubit quantum secret sharing},
  author={Schmid, Christian and Trojek, Pavel and Bourennane, Mohamed and Kurtsiefer, Christian and {\.Z}ukowski, Marek and Weinfurter, Harald},
  journal={Physical review letters},
  volume={95},
  number={23},
  pages={230505},
  year={2005},
  publisher={APS}
}

@article{harn2017share,
  title={How to share secret efficiently over networks},
  author={Harn, Lein and Hsu, Ching-Fang and Xia, Zhe and Zhou, Junwei},
  journal={Security and Communication Networks},
  volume={2017},
  number={1},
  pages={5437403},
  year={2017},
  publisher={Wiley Online Library}
}

@article{tittel2001experimental,
  title={Experimental demonstration of quantum secret sharing},
  author={Tittel, Wolfgang and Zbinden, Hugo and Gisin, Nicolas},
  journal={Physical Review A},
  volume={63},
  number={4},
  pages={042301},
  year={2001},
  publisher={APS}
}

@article{verstraete2002four,
  title={Four qubits can be entangled in nine different ways},
  author={Verstraete, Frank and Dehaene, Jeroen and De Moor, Bart and Verschelde, Henri},
  journal={Physical Review A},
  volume={65},
  number={5},
  pages={052112},
  year={2002},
  publisher={APS}
}

@article{keet2010quantum,
  title={Quantum secret sharing with qudit graph states},
  author={Keet, Adrian and Fortescue, Ben and Markham, Damian and Sanders, Barry C},
  journal={Physical Review A},
  volume={82},
  number={6},
  pages={062315},
  year={2010},
  publisher={APS}
}

@article{ray2016sequential,
  title={Sequential quantum secret sharing in a noisy environment aided with weak measurements},
  author={Ray, Maharshi and Chatterjee, Sourav and Chakrabarty, Indranil},
  journal={The European Physical Journal D},
  volume={70},
  pages={1--11},
  year={2016},
  publisher={Springer}
}

@article{chakrabarty2011deletion,
  title={Deletion, Bell’s inequality, teleportation},
  author={Chakrabarty, Indranil and Ganguly, Nirman and Choudhury, Binayak S},
  journal={Quantum Information Processing},
  volume={10},
  number={1},
  pages={27--32},
  year={2011},
  publisher={Springer}
}

@Inbook{Blakley2011,
author="Blakley, G. R.
and Kabatiansky, Gregory",
editor="van Tilborg, Henk C. A.
and Jajodia, Sushil",
title="Shamir's Threshold Scheme",
bookTitle="Encyclopedia of Cryptography and Security",
year="2011",
publisher="Springer US",
address="Boston, MA",
pages="1193--1194",
isbn="978-1-4419-5906-5",
doi="10.1007/978-1-4419-5906-5_390",
url="https://doi.org/10.1007/978-1-4419-5906-5_390"
}

@article{10.1145/1008908.1008920,
author = {Wiesner, Stephen},
title = {Conjugate coding},
year = {1983},
issue_date = {Winter-Spring 1983},
publisher = {Association for Computing Machinery},
address = {New York, NY, USA},
volume = {15},
number = {1},
issn = {0163-5700},
url = {https://doi.org/10.1145/1008908.1008920},
doi = {10.1145/1008908.1008920},
journal = {SIGACT News},
month = jan,
pages = {78–88},
numpages = {11}
}

@article{Gottesman_2000,
   title={Theory of quantum secret sharing},
   volume={61},
   ISSN={1094-1622},
   url={http://dx.doi.org/10.1103/PhysRevA.61.042311},
   DOI={10.1103/physreva.61.042311},
   number={4},
   journal={Physical Review A},
   publisher={American Physical Society (APS)},
   author={Gottesman, Daniel},
   year={2000},
   month=mar }

@article{ refId0,
	author = {{Wang, Z.-Y.} and {Yuan, H.} and {Shi, S.-H.} and {Zhang, Z.-J.}},
	title = {Three-party qutrit-state sharing},
	DOI= "10.1140/epjd/e2006-00215-y",
	url= "https://doi.org/10.1140/epjd/e2006-00215-y",
	journal = {Eur. Phys. J. D},
	year = 2007,
	volume = 41,
	number = 2,
	pages = "371-375",
	month = "",
}
\begin{widetext}
\noindent\textit{\bf Appendix 1:} (Proof of Theorem 1)\\
\\
For Alice to successfully retrieve the state, Bob must measures his qubit and Charlie must measure his respective qubit in the Hadamard Basis. We show how Alice can retrieve back her message by enlisting down the respective local operations corresponding to Bob’s and Charlie’s measurement outcomes. \\ \\
If Bob measures measures his qubit in Hadamard Basis $\lvert+\rangle$ and Charlie  measures measures his respective qubit in Hadamard Basis $\lvert+\rangle$, the resultant state on Alice’s side is given by the following (when Alice's initial measurement outcome is $\lvert\phi^+\rangle$):
\begin{eqnarray}
\sqrt{2}\{(a\alpha + b\beta)|0\rangle + (b\alpha + a\beta)|1\rangle\} {}\nonumber
\end{eqnarray}
Condition of normalization of the above Alice's resultant state is as follows:
\begin{eqnarray}
|a|^2+|b|^2+(\tilde{a}b+a\tilde{b})(\alpha\tilde\beta+\tilde\alpha\beta)
 =  \frac{1}{2} {}\nonumber&&  
\end{eqnarray}
In order to find Alice's Local Operations state for retrieving the state, we proceed by taking a Unitary Matrix as follows:
\begin{equation*}
U = 
\begin{bmatrix}
a_{1} + ib_{1} & c_{1} - id_{1} \\
c_{1} + id_{1} & -a_{1} + ib_{1} \\
\end{bmatrix}
\end{equation*}
such that det(U) = -1
\\
\\
We need to find $a_{1}, b_{1}, c_{1}, d_{1}$ by the solving the following equation for Alice to retrieve the state:

\[
\begin{bmatrix}
    a_{1} + ib_{1} & c_{1} - id_{1} \\
    c_{1} + id_{1} & -a_{1} + ib_{1}      \\
          
\end{bmatrix}
\begin{bmatrix}
    \sqrt{2}(a\alpha + b\beta) \\
    \sqrt{2}(b\alpha + a\beta)    \\
          
\end{bmatrix}
= 
\begin{bmatrix}
    \alpha      \\
    \beta      
\end{bmatrix} 
\]
\\
As $a, b$, $\alpha$ and $\beta$ are complex numbers, we write $a = x_{1} + iy_{1}, b = x_{2} + iy_{2}, \alpha = x_{3} + iy_{3}, \beta = x_{4} + iy_{4}$. 
\\
The above equation transforms as follows:
\[
\begin{bmatrix}
    a_{1} + ib_{1} & c_{1} - id_{1} \\
    c_{1} + id_{1} & -a_{1} + ib_{1}      \\
          
\end{bmatrix}
\begin{bmatrix}
    \sqrt{2}(\lambda_{1} + i\lambda_{2}) \\
    \sqrt{2}(\lambda_{3} + i\lambda_{4})    \\
          
\end{bmatrix}
= 
\begin{bmatrix}
    \alpha      \\
    \beta      
\end{bmatrix} 
\]
\\
Where we have:\\
\begin{eqnarray}
\lambda_{1} = x_{1}x_{3} + x_{2}x_{4} - y_{1}y_{3} - y_{2}y_{4}{}\nonumber\\
\lambda_{2} = x_{3}y_{1} + x_{1}y_{3} + x_{2}y_{4} + x_{4}y_{2}{}\nonumber\\
\lambda_{3} = x_{2}x_{3} + x_{1}x_{4} - y_{1}y_{4} - y_{2}y_{3}{}\nonumber\\
\lambda_{4} = x_{3}y_{2} + x_{4}y_{1} + x_{2}y_{3} + x_{1}y_{4}{}\nonumber
\end{eqnarray}
\\
Such that:\\
\begin{eqnarray}
\lambda_{1}^2 + \lambda_{2}^2 + \lambda_{3}^2 + \lambda_{4}^2 = |a|^2+|b|^2+(\tilde{a}b+a\tilde{b})(\alpha\tilde\beta+\tilde\alpha\beta) =  \frac{1}{2}
\end{eqnarray}
Solving the above equation by comparing real and imaginary parts on both sides to get:
\begin{eqnarray}
a_{1} = \frac{\lambda_{1}x_{3} + \lambda_{2}y_{3} - \lambda_{3}x_{4} - \lambda_{4}y{4}}{\sqrt{2}{(\lambda_{1}^2 + \lambda_{2}^2 + \lambda_{3}^2 + \lambda_{4}^2)}} = \sqrt{2}(\lambda_{1}x_{3} + \lambda_{2}y_{3} - \lambda_{3}x_{4} - \lambda_{4}y_{4}) {}\nonumber\\
b_{1} = \frac{\lambda_{1}y_{3} - \lambda_{2}x_{3} + \lambda_{3}y_{4} - \lambda_{4}x_{4}}{\sqrt{2}{(\lambda_{1}^2 + \lambda_{2}^2 + \lambda_{3}^2 + \lambda_{4}^2)}} = \sqrt{2}(\lambda_{1}y_{3} - \lambda_{2}x_{3} + \lambda_{3}y_{4} - \lambda_{4}x_{4}) {}\nonumber\\
c_{1} = \frac{\lambda_{1}x_{4} + \lambda_{2}y_{4} + \lambda_{3}x_{3} + \lambda_{4}y_{3}}{\sqrt{2}{(\lambda_{1}^2 + \lambda_{2}^2 + \lambda_{3}^2 + \lambda_{4}^2)}} = \sqrt{2}(\lambda_{1}x_{4} + \lambda_{2}y_{4} + \lambda_{3}x_{3} + \lambda_{4}y_{3}) {}\nonumber\\
d_{1} = \frac{\lambda_{4}x_{3} - \lambda_{3}y_{3} + \lambda_{1}y_{4} - \lambda_{2}x_{4}}{\sqrt{2}{(\lambda_{1}^2 + \lambda_{2}^2 + \lambda_{3}^2 + \lambda_{4}^2)}} = \sqrt{2}(\lambda_{4}x_{3} - \lambda_{3}y_{3} + \lambda_{1}y_{4} - \lambda_{2}x_{4}) {}\nonumber
\end{eqnarray}
\begin{eqnarray}
a_{1} + ib_{1} = \sqrt{2}\{(\lambda_{1} - i\lambda_{2})\alpha - (\lambda_{3} + i\lambda_{4})\tilde{\beta}\} {}\nonumber\\
= \sqrt{2}\{(\tilde{a}\tilde{\alpha} + \tilde{b}\tilde{\beta})\alpha - (b\alpha + a\beta)\tilde{\beta}\} {}\nonumber\\
= \sqrt{2}\{(\tilde{a} - (\tilde{a} + a)|\beta|^2 + (\tilde{b} - b)\alpha\tilde{\beta})\} 
\end{eqnarray}
\begin{eqnarray}
-a_{1} + ib_{1} = \sqrt{2}\{-(\lambda_{1} + i\lambda_{2})\tilde{\alpha} + (\lambda_{3} - i\lambda_{4})\beta\} {}\nonumber\\
= \sqrt{2}\{-(a\alpha + b\beta)\tilde{\alpha} + (\tilde{a}\tilde{\beta} + \tilde{b}\tilde{\alpha})\beta\} {}\nonumber\\
= \sqrt{2}\{-a + (\tilde{a} + a)|\beta|^2 + (\tilde{b} - b)\tilde{\alpha}\beta\} 
\end{eqnarray}
\begin{eqnarray}
c_{1} + id_{1} = \sqrt{2}\{(\lambda_{1} - i\lambda_{2})\beta + (\lambda_{3} + i\lambda_{4})\tilde{\alpha}\} {}\nonumber\\
= \sqrt{2}\{(\tilde{a}\tilde{\alpha} + \tilde{b}\tilde{\beta})\beta + (b\alpha + a\beta)\tilde{\alpha}\} {}\nonumber\\
= \sqrt{2}\{\tilde{b} + (\tilde{a} + a)\tilde{\alpha}\beta - (\tilde{b} - b)|\alpha|^2\}
\end{eqnarray}
\begin{eqnarray}
c_{1} - id_{1} = \sqrt{2}\{(\lambda_{1} + i\lambda_{2})\tilde{\beta} + (\lambda_{3} - i\lambda_{4})\alpha\} {}\nonumber\\
= \sqrt{2}\{(a\alpha + b\beta)\tilde{\beta} + (\tilde{a}\tilde{\beta} + \tilde{b}\tilde{\alpha})\alpha\} {}\nonumber\\
= \sqrt{2}\{b + (\tilde{a} + a)\alpha\tilde{\beta} + (\tilde{b} - b)|\alpha|^2\}
\end{eqnarray}
Using equations (9), (10), (11) and (12), we obtain the Unitary matrix as follows:\\
\begin{equation*}
\begin{bmatrix}
    \sqrt{2}\{(\tilde{a} - (\tilde{a} + a)|\beta|^2 + (\tilde{b} - b)\alpha\tilde{\beta})\} & \sqrt{2}\{b + (\tilde{a} + a)\alpha\tilde{\beta} + (\tilde{b} - b)|\alpha|^2\} \\
    \sqrt{2}\{\tilde{b} + (\tilde{a} + a)\tilde{\alpha}\beta - (\tilde{b} - b)|\alpha|^2\} & \sqrt{2}\{-a + (\tilde{a} + a)|\beta|^2 + (\tilde{b} - b)\tilde{\alpha}\beta\}      \\
          
\end{bmatrix}
\end{equation*}
\\
\\
Since the elements of the matrix need to be independent of $\alpha$ and $\beta$, we have the following conditions:\\
\begin{eqnarray}
\tilde{a} + a = 0
\end{eqnarray}
\begin{eqnarray}
\tilde{b} - b = 0
\end{eqnarray}
From eq (13) we have $a$ to be purely imaginary and from eq (14) we have $b$ to be purely real.
After the imposition of the above conditions eq (13) and (14), the Unitary Matrix reduces to:\\
\begin{equation*}
U =
\begin{bmatrix}
    \sqrt{2}\tilde{a} & \sqrt{2}b \\
    \sqrt{2}\tilde{b} & -\sqrt{2}a      \\
          
\end{bmatrix}
= \begin{bmatrix}
    -\sqrt{2}a & \sqrt{2}b \\
    \sqrt{2}b & -\sqrt{2}a      \\
          
\end{bmatrix}
\end{equation*}
Where we made use of eq (14) and eq (15) above, $\tilde{a} + a = 0$ or $\tilde{a} = -a$ and $\tilde{b} - b = 0$ or $\tilde{b} = b$
\\
\\
det(U) = $2a^2 - 2b^2 = -1$ or $2b^2 - 2a^2 = 1$\\
\\
Using the above matrix, Alice successfully retrieves the state as follows:
\[
\begin{bmatrix}
    -\sqrt{2}a & \sqrt{2}b \\
    \sqrt{2}b & -\sqrt{2}a      \\
          
\end{bmatrix}
\begin{bmatrix}
    \sqrt{2}(a\alpha + b\beta) \\
    \sqrt{2}(b\alpha + a\beta)    \\
          
\end{bmatrix}
= 
\begin{bmatrix}
    2(-a^2\alpha - ab\beta + b^2\alpha + ab\beta)      \\
    2(ab\alpha + b^2\beta -ab\alpha - a^2\beta )    
\end{bmatrix} 
=
\begin{bmatrix}
    2(b^2 - a^2)\alpha      \\
    2(b^2 - a^2)\beta      
\end{bmatrix} 
=
\begin{bmatrix}
    \alpha      \\
    \beta      
\end{bmatrix} 
\]
Expressing the above matrix in terms of Pauli Matrices, we have U = $\sqrt{2}(-aI + b\sigma_x)$.
Now as $2b^2 - 2a^2 = 1$, we have $a^2 = b^2 - \frac{1}{2}$
\\
\\
Since $a$ is purely imaginary, we must have $b^2 \leq \frac{1}{2}$ or $-\frac{1}{\sqrt{2}} \leq b \leq \frac{1}{\sqrt{2}}$.
Now from the condition of normalization of Alice's resultant state, we have:
\begin{eqnarray}
|a|^2+|b|^2+(\tilde{a}b+a\tilde{b})(\alpha\tilde\beta+\tilde\alpha\beta)
 =  \frac{1}{2} {}\nonumber&&  
\end{eqnarray}
As $b = \tilde{b}$, the above equation reduces to: 
\begin{eqnarray}
|a|^2+|b|^2+b(\tilde{a}+a)(\alpha\tilde\beta+\tilde\alpha\beta)
 =  \frac{1}{2} {}\nonumber   
\end{eqnarray}
And as $\tilde{a}+a$ = 0, the above equation further reduces to:
\begin{eqnarray}
|a|^2+|b|^2 = \frac{1}{2} {}\nonumber&&  
\end{eqnarray}
As $b$ is a real number, $b^2 = |b|^2$ and so we have:
\begin{eqnarray}
|a|^2+b^2 =  \frac{1}{2}{}\nonumber &&  
\end{eqnarray}
or
\begin{eqnarray}
|a|^2 =  \frac{1}{2} - b^2 {}\nonumber&&  
\end{eqnarray}
\\

\noindent\textit{\bf Appendix 2:} (Proof of Theorem 1)\\
\\
For Alice to successfully retrieve the state, Bob must measures his qubit and Charlie must measure his respective qubit in the Hadamard Basis. We show how Alice can retrieve back her message by enlisting down the respective local operations corresponding to Bob’s and Charlie’s measurement outcomes.\\ \\
If Bob measures his qubit in Hadamard Basis $\lvert+\rangle$  and Charlie measures his respective qubit in Hadamard Basis $\lvert-\rangle$, the resultant state on Alice’s side is given by the following (when Alice's initial measurement outcome is $\lvert\phi^+\rangle$):
\begin{eqnarray}
\sqrt{2}\{(d\alpha + c\beta)|0\rangle - (c\alpha + d\beta)|1\rangle\} {}\nonumber
\end{eqnarray}
Condition of normalization of the above Alice's resultant state is as follows:
\begin{eqnarray}
|c|^2+|d|^2+(\tilde{c}d+d\tilde{c})(\alpha\tilde\beta+\tilde\alpha\beta)
 =  \frac{1}{2} {}\nonumber&&  
\end{eqnarray}
In order to find Alice's Local Operations state for retrieving the state, we proceed by taking a Unitary Matrix as follows:
\begin{equation*}
U = 
\begin{bmatrix}
a_{1} + ib_{1} & c_{1} - id_{1} \\
c_{1} + id_{1} & -a_{1} + ib_{1} \\
\end{bmatrix}
\end{equation*}
such that det(U) = -1
\\
\\
We need to find $a_{1}, b_{1}, c_{1}, d_{1}$ by the solving the following equation for Alice to retrieve the state:

\[
\begin{bmatrix}
    a_{1} + ib_{1} & c_{1} - id_{1} \\
    c_{1} + id_{1} & -a_{1} + ib_{1}      \\
          
\end{bmatrix}
\begin{bmatrix}
    \sqrt{2}(d\alpha + c\beta) \\
    -\sqrt{2}(c\alpha + d\beta)    \\
          
\end{bmatrix}
= 
\begin{bmatrix}
    \alpha      \\
    \beta      
\end{bmatrix} 
\]
\\
As $a, b$, $\alpha$ and $\beta$ are complex numbers, we write $d = x_{1} + iy_{1}, c = x_{2} + iy_{2}, \alpha = x_{3} + iy_{3}, \beta = x_{4} + iy_{4}$. 
\\
The above equation transforms as follows:
\[
\begin{bmatrix}
    a_{1} + ib_{1} & c_{1} - id_{1} \\
    c_{1} + id_{1} & -a_{1} + ib_{1}      \\
          
\end{bmatrix}
\begin{bmatrix}
    \sqrt{2}(\lambda_{1} + i\lambda_{2}) \\
    -\sqrt{2}(\lambda_{3} + i\lambda_{4})    \\
          
\end{bmatrix}
= 
\begin{bmatrix}
    \alpha      \\
    \beta      
\end{bmatrix} 
\]
\\
Where we have:\\
\begin{eqnarray}
\lambda_{1} = x_{1}x_{3} + x_{2}x_{4} - y_{1}y_{3} - y_{2}y_{4}{}\nonumber\\
\lambda_{2} = x_{3}y_{1} + x_{1}y_{3} + x_{2}y_{4} + x_{4}y_{2}{}\nonumber\\
\lambda_{3} = x_{2}x_{3} + x_{1}x_{4} - y_{1}y_{4} - y_{2}y_{3}{}\nonumber\\
\lambda_{4} = x_{3}y_{2} + x_{4}y_{1} + x_{2}y_{3} + x_{1}y_{4}{}\nonumber
\end{eqnarray}
\\
Such that:\\
\begin{eqnarray}
\lambda_{1}^2 + \lambda_{2}^2 + \lambda_{3}^2 + \lambda_{4}^2 = |a|^2+|b|^2+(\tilde{a}b+a\tilde{b})(\alpha\tilde\beta+\tilde\alpha\beta) =  \frac{1}{2}
\end{eqnarray}
Solving the above equation by comparing real and imaginary parts on both sides to get:
\begin{eqnarray}
a_{1} = \frac{\lambda_{1}x_{3} + \lambda_{2}y_{3} + \lambda_{3}x_{4} + \lambda_{4}y{4}}{\sqrt{2}{(\lambda_{1}^2 + \lambda_{2}^2 + \lambda_{3}^2 + \lambda_{4}^2)}} = \sqrt{2}(\lambda_{1}x_{3} + \lambda_{2}y_{3} + \lambda_{3}x_{4} + \lambda_{4}y_{4}) {}\nonumber\\
b_{1} = \frac{\lambda_{1}y_{3} - \lambda_{2}x_{3} - \lambda_{3}y_{4} + \lambda_{4}x_{4}}{\sqrt{2}{(\lambda_{1}^2 + \lambda_{2}^2 + \lambda_{3}^2 + \lambda_{4}^2)}} = \sqrt{2}(\lambda_{1}y_{3} - \lambda_{2}x_{3} - \lambda_{3}y_{4} + \lambda_{4}x_{4}) {}\nonumber\\
c_{1} = \frac{\lambda_{1}x_{4} + \lambda_{2}y_{4} - \lambda_{3}x_{3} - \lambda_{4}y_{3}}{\sqrt{2}{(\lambda_{1}^2 + \lambda_{2}^2 + \lambda_{3}^2 + \lambda_{4}^2)}} = \sqrt{2}(\lambda_{1}x_{4} + \lambda_{2}y_{4} - \lambda_{3}x_{3} - \lambda_{4}y_{3}) {}\nonumber\\
d_{1} = \frac{-\lambda_{4}x_{3} + \lambda_{3}y_{3} + \lambda_{1}y_{4} - \lambda_{2}x_{4}}{\sqrt{2}{(\lambda_{1}^2 + \lambda_{2}^2 + \lambda_{3}^2 + \lambda_{4}^2)}} = \sqrt{2}(-\lambda_{4}x_{3} + \lambda_{3}y_{3} + \lambda_{1}y_{4} - \lambda_{2}x_{4}) {}\nonumber
\end{eqnarray}
\begin{eqnarray}
a_{1} + ib_{1} = \sqrt{2}\{(\lambda_{1} - i\lambda_{2})\alpha + (\lambda_{3} + i\lambda_{4})\tilde{\beta}\} {}\nonumber\\
= \sqrt{2}\{(\tilde{d}\tilde{\alpha} + \tilde{c}\tilde{\beta})\alpha + (c\alpha + d\beta)\tilde{\beta}\} {}\nonumber\\
= \sqrt{2}\{(\tilde{d} + (d - \tilde{d})|\beta|^2 + (c + \tilde{c})\alpha\tilde{\beta})\} 
\end{eqnarray}
\begin{eqnarray}
-a_{1} + ib_{1} = \sqrt{2}\{-(\lambda_{1} + i\lambda_{2})\tilde{\alpha} - (\lambda_{3} - i\lambda_{4})\beta\} {}\nonumber\\
= \sqrt{2}\{-(d\alpha + c\beta)\tilde{\alpha} - (\tilde{d}\tilde{\beta} + \tilde{c}\tilde{\alpha})\beta\} {}\nonumber\\
= \sqrt{2}\{-d + (d - \tilde{d})|\beta|^2 - (c + \tilde{c})\tilde{\alpha}\beta\} 
\end{eqnarray}
\begin{eqnarray}
c_{1} + id_{1} = \sqrt{2}\{(\lambda_{1} - i\lambda_{2})\beta - (\lambda_{3} + i\lambda_{4})\tilde{\alpha}\} {}\nonumber\\
= \sqrt{2}\{(\tilde{d}\tilde{\alpha} + \tilde{c}\tilde{\beta})\beta - (c\alpha + d\beta)\tilde{\alpha}\} {}\nonumber\\
= \sqrt{2}\{\tilde{c} + (\tilde{d} - d)\tilde{\alpha}\beta + (\tilde{c} + c)|\alpha|^2\}
\end{eqnarray}
\begin{eqnarray}
c_{1} - id_{1} = \sqrt{2}\{(\lambda_{1} + i\lambda_{2})\tilde{\beta} - (\lambda_{3} - i\lambda_{4})\alpha\} {}\nonumber\\
= \sqrt{2}\{(d\alpha + c\beta)\tilde{\beta} - (\tilde{d}\tilde{\beta} + \tilde{c}\tilde{\alpha})\alpha\} {}\nonumber\\
= \sqrt{2}\{c - (d - \tilde{d})\alpha\tilde{\beta} - (\tilde{c} + c)|\alpha|^2\}
\end{eqnarray}
Using equations (16), (17), (18) and (19), we obtain the Unitary matrix as follows:\\
\begin{equation*}
\begin{bmatrix}
    \sqrt{2}\{(\tilde{d} + (d - \tilde{d})|\beta|^2 + (c + \tilde{c})\alpha\tilde{\beta})\} & \sqrt{2}\{c - (d - \tilde{d})\alpha\tilde{\beta} - (\tilde{c} + c)|\alpha|^2\} \\
    \sqrt{2}\{\tilde{c} + (\tilde{d} - d)\tilde{\alpha}\beta + (\tilde{c} + c)|\alpha|^2\} & \sqrt{2}\{-d + (d - \tilde{d})|\beta|^2 - (c + \tilde{c})\tilde{\alpha}\beta\}      \\
          
\end{bmatrix}
\end{equation*}
\\
\\
Since the elements of the matrix need to be independent of $\alpha$ and $\beta$, we have the following conditions:\\
\begin{eqnarray}
d - \tilde{d} = 0
\end{eqnarray}
\begin{eqnarray}
c + \tilde{c} = 0
\end{eqnarray}
From eq (20) we have $d$ to be purely real and from eq (21) we have $c$ to be purely imaginary.
After the imposition of the above conditions eq (20) and (21), the Unitary Matrix reduces to:\\
\begin{equation*}
U =
\begin{bmatrix}
    \sqrt{2}\tilde{d} & \sqrt{2}c \\
    \sqrt{2}\tilde{c} & -\sqrt{2}d      \\
          
\end{bmatrix}
= \begin{bmatrix}
    \sqrt{2}d & \sqrt{2}c \\
    -\sqrt{2}c & -\sqrt{2}d      \\
   
   \end{bmatrix}
\end{equation*}
Where we made use of eq (20) and eq (21) above, $d - \tilde{d} = 0$ or $d = \tilde{d}$ and $c + \tilde{c} = 0$ or $c = -\tilde{c}$
\\
\\
det(U) = $2c^2 - 2d^2 = -1$ or $2d^2 - 2c^2 = 1$\\
\\
Using the above matrix, Alice successfully retrieves the state as follows:
\[
\begin{bmatrix}
    \sqrt{2}d & \sqrt{2}c \\
    -\sqrt{2}c & -\sqrt{2}d      \\
\end{bmatrix}
\begin{bmatrix}
    \sqrt{2}(d\alpha + c\beta) \\
    -\sqrt{2}(c\alpha + d\beta)    \\
   
\end{bmatrix}
= 
\begin{bmatrix}
    2(d^2\alpha + cd\beta - c^2\alpha - cd\beta)      \\
    2(-cd\alpha - c^2\beta + cd\alpha + d^2\beta )    
\end{bmatrix} 
=
\begin{bmatrix}
    2(d^2 - c^2)\alpha      \\
    2(d^2 - c^2)\beta      
\end{bmatrix} 
=
\begin{bmatrix}
    \alpha      \\
    \beta      
\end{bmatrix} 
\]
\\
Expressing the above matrix in terms of Pauli Matrices, we have U = $\sqrt{2}(d\sigma_x + ic\sigma_y)$
\\
\\
Now as $2d^2 - 2c^2 = 1$, we have $c^2 = d^2 - \frac{1}{2}$
\\
\\
Since $c$ is purely imaginary, we must have $d^2 \leq \frac{1}{2}$ or $-\frac{1}{\sqrt{2}} \leq d \leq \frac{1}{\sqrt{2}}$
Now from the condition of normalization of Alice's resultant state, we have:
\begin{eqnarray}
|d|^2+|c|^2+(\tilde{d}c+d\tilde{c})(\alpha\tilde\beta+\tilde\alpha\beta)
 =  \frac{1}{2} {}\nonumber&&  
\end{eqnarray}
As $d = \tilde{d}$, the above equation reduces to: 
\begin{eqnarray}
|d|^2+|c|^2+d(\tilde{c}+c)(\alpha\tilde\beta+\tilde\alpha\beta)
 =  \frac{1}{2} {}\nonumber   
\end{eqnarray}
And as $\tilde{c}+c$ = 0, the above equation further reduces to:
\begin{eqnarray}
|d|^2+|c|^2 = \frac{1}{2} {}\nonumber&&  
\end{eqnarray}
As $d$ is a real number, $d^2 = |d|^2$ and so we have:
\begin{eqnarray}
|c|^2+d^2 =  \frac{1}{2}{}\nonumber &&  
\end{eqnarray}
or
\begin{eqnarray}
|c|^2 =  \frac{1}{2} - d^2 {}\nonumber&&  
\end{eqnarray}
\\

\noindent\textit{\bf Appendix 3:} (Proof of Theorem 1)\\ \\
For Alice to successfully retrieve the secret, Bob must measures his qubit and Charlie must measure his respective qubit in the Hadamard Basis. We show how Alice can retrieve back her message by enlisting down the respective local operations corresponding to Bob’s and Charlie’s measurement outcomes. \\ \\
If Bob measures his qubit in Hadamard Basis $\lvert+\rangle$ and Charlie measures his respective qubit in Hadamard Basis $\lvert-\rangle$, the resultant state on Alice’s side is given by the following (when Alice's initial measurement outcome is $\lvert\phi^-\rangle$):
\begin{eqnarray}
\sqrt{2}\{(d\alpha - c\beta)|0\rangle -
(c\alpha - d\beta)|1\rangle\} {}\nonumber
\end{eqnarray}
Condition of normalization of the above Alice's resultant state is as follows:
\begin{eqnarray}
|c|^2+|d|^2+(\tilde{c}d+d\tilde{c})(\alpha\tilde\beta+\tilde\alpha\beta)
 =  \frac{1}{2} {}\nonumber&&  
\end{eqnarray}
In order to find Alice's Local Operations state for retrieving the state, we proceed by taking a Unitary Matrix as follows:
\begin{equation*}
U = 
\begin{bmatrix}
a_{1} + ib_{1} & c_{1} - id_{1} \\
c_{1} + id_{1} & -a_{1} + ib_{1} \\
\end{bmatrix}
\end{equation*}
such that det(U) = -1
\\
\\
We need to find $a_{1}, b_{1}, c_{1}, d_{1}$ by the solving the following equation for Alice to retrieve the state:

\[
\begin{bmatrix}
    a_{1} + ib_{1} & c_{1} - id_{1} \\
    c_{1} + id_{1} & -a_{1} + ib_{1}      \\
          
\end{bmatrix}
\begin{bmatrix}
    \sqrt{2}(d\alpha - c\beta) \\
    -\sqrt{2}(c\alpha - d\beta)    \\
          
\end{bmatrix}
= 
\begin{bmatrix}
    \alpha      \\
    \beta      
\end{bmatrix} 
\]
\\
As $a, b$, $\alpha$ and $\beta$ are complex numbers, we write $d = x_{1} + iy_{1}, c = x_{2} + iy_{2}, \alpha = x_{3} + iy_{3}, \beta = x_{4} + iy_{4}$. 
\\
The above equation transforms as follows:
\[
\begin{bmatrix}
    a_{1} + ib_{1} & c_{1} - id_{1} \\
    c_{1} + id_{1} & -a_{1} + ib_{1}      \\
          
\end{bmatrix}
\begin{bmatrix}
    \sqrt{2}(\lambda_{1} + i\lambda_{2}) \\
    \sqrt{2}(\lambda_{3} + i\lambda_{4})    \\
          
\end{bmatrix}
= 
\begin{bmatrix}
    \alpha      \\
    \beta      
\end{bmatrix} 
\]
\\
Where we have:\\
\begin{eqnarray}
\lambda_{1} = x_{1}x_{3} - x_{2}x_{4} - y_{1}y_{3} + y_{2}y_{4}{}\nonumber\\
\lambda_{2} = x_{3}y_{1} + x_{1}y_{3} - x_{2}y_{4} - x_{4}y_{2}{}\nonumber\\
\lambda_{3} = -x_{2}x_{3} + x_{1}x_{4} - y_{1}y_{4} + y_{2}y_{3}{}\nonumber\\
\lambda_{4} = -x_{3}y_{2} + x_{4}y_{1} - x_{2}y_{3} + x_{1}y_{4}{}\nonumber
\end{eqnarray}
\\
Such that:\\
\begin{eqnarray}
\lambda_{1}^2 + \lambda_{2}^2 + \lambda_{3}^2 + \lambda_{4}^2 = |d|^2+|c|^2+(\tilde{d}c+d\tilde{c})(\alpha\tilde\beta+\tilde\alpha\beta) =  \frac{1}{2}
\end{eqnarray}
Solving the above equation by comparing real and imaginary parts on both sides to get:
\begin{eqnarray}
a_{1} = \frac{\lambda_{1}x_{3} + \lambda_{2}y_{3} - \lambda_{3}x_{4} - \lambda_{4}y{4}}{\sqrt{2}{(\lambda_{1}^2 + \lambda_{2}^2 + \lambda_{3}^2 + \lambda_{4}^2)}} = \sqrt{2}(\lambda_{1}x_{3} + \lambda_{2}y_{3} - \lambda_{3}x_{4} - \lambda_{4}y_{4}) {}\nonumber\\
b_{1} = \frac{\lambda_{1}y_{3} - \lambda_{2}x_{3} + \lambda_{3}y_{4} - \lambda_{4}x_{4}}{\sqrt{2}{(\lambda_{1}^2 + \lambda_{2}^2 + \lambda_{3}^2 + \lambda_{4}^2)}} = \sqrt{2}(\lambda_{1}y_{3} - \lambda_{2}x_{3} + \lambda_{3}y_{4} - \lambda_{4}x_{4}) {}\nonumber\\
c_{1} = \frac{\lambda_{1}x_{4} + \lambda_{2}y_{4} + \lambda_{3}x_{3} + \lambda_{4}y_{3}}{\sqrt{2}{(\lambda_{1}^2 + \lambda_{2}^2 + \lambda_{3}^2 + \lambda_{4}^2)}} = \sqrt{2}(\lambda_{1}x_{4} + \lambda_{2}y_{4} + \lambda_{3}x_{3} + \lambda_{4}y_{3}) {}\nonumber\\
d_{1} = \frac{\lambda_{4}x_{3} - \lambda_{3}y_{3} + \lambda_{1}y_{4} - \lambda_{2}x_{4}}{\sqrt{2}{(\lambda_{1}^2 + \lambda_{2}^2 + \lambda_{3}^2 + \lambda_{4}^2)}} = \sqrt{2}(\lambda_{4}x_{3} - \lambda_{3}y_{3} + \lambda_{1}y_{4} - \lambda_{2}x_{4}) {}\nonumber
\end{eqnarray}
\begin{eqnarray}
a_{1} + ib_{1} = \sqrt{2}\{(\lambda_{1} - i\lambda_{2})\alpha - (\lambda_{3} + i\lambda_{4})\tilde{\beta}\} {}\nonumber\\
= \sqrt{2}\{(\tilde{d}\tilde{\alpha} - \tilde{c}\tilde{\beta})\alpha - (d\beta - c\alpha)\tilde{\beta}\} {}\nonumber\\
= \sqrt{2}\{(\tilde{d} - (d + \tilde{d})|\beta|^2 + (c - \tilde{c})\alpha\tilde{\beta})\} 
\end{eqnarray}
\begin{eqnarray}
-a_{1} + ib_{1} = \sqrt{2}\{-(\lambda_{1} + i\lambda_{2})\tilde{\alpha} + (\lambda_{3} - i\lambda_{4})\beta\} {}\nonumber\\
= \sqrt{2}\{-(d\alpha - c\beta)\tilde{\alpha} + (\tilde{d}\tilde{\beta} - \tilde{c}\tilde{\alpha})\beta\} {}\nonumber\\
= \sqrt{2}\{-d + (d + \tilde{d})|\beta|^2 + (c - \tilde{c})\tilde{\alpha}\beta\} 
\end{eqnarray}
\begin{eqnarray}
c_{1} + id_{1} = \sqrt{2}\{(\lambda_{1} - i\lambda_{2})\beta + (\lambda_{3} + i\lambda_{4})\tilde{\alpha}\} {}\nonumber\\
= \sqrt{2}\{(\tilde{d}\tilde{\alpha} - \tilde{c}\tilde{\beta})\beta + (d\beta - c\alpha)\tilde{\alpha}\} {}\nonumber\\
= \sqrt{2}\{-\tilde{c} + (\tilde{d} + d)\tilde{\alpha}\beta + (\tilde{c} - c)|\alpha|^2\}
\end{eqnarray}
\begin{eqnarray}
c_{1} - id_{1} = \sqrt{2}\{(\lambda_{1} + i\lambda_{2})\tilde{\beta} + (\lambda_{3} - i\lambda_{4})\alpha\} {}\nonumber\\
= \sqrt{2}\{(d\alpha - c\beta)\tilde{\beta} + (\tilde{d}\tilde{\beta} - \tilde{c}\tilde{\alpha})\alpha\} {}\nonumber\\
= \sqrt{2}\{-c + (d + \tilde{d})\alpha\tilde{\beta} + (c - \tilde{c})|\alpha|^2\}
\end{eqnarray}
Using equations (23), (24), (25) and (26), we obtain the Unitary matrix as follows:\\
\begin{equation*}
\begin{bmatrix}
    \sqrt{2}\{(\tilde{d} - (d + \tilde{d})|\beta|^2 + (c - \tilde{c})\alpha\tilde{\beta})\} & \sqrt{2}\{-c + (d + \tilde{d})\alpha\tilde{\beta} + (c - \tilde{c})|\alpha|^2\} \\
    \sqrt{2}\{-\tilde{c} + (\tilde{d} + d)\tilde{\alpha}\beta + (\tilde{c} - c)|\alpha|^2\} & \sqrt{2}\{-d + (d + \tilde{d})|\beta|^2 + (c - \tilde{c})\tilde{\alpha}\beta\}     \\
          
\end{bmatrix}
\end{equation*}
\\
\\
Since the elements of the matrix need to be independent of $\alpha$ and $\beta$, we have the following conditions:\\
\begin{eqnarray}
d + \tilde{d} = 0
\end{eqnarray}
\begin{eqnarray}
c - \tilde{c} = 0
\end{eqnarray}
From eq (27) we have $d$ to be purely imaginary and from eq (28) we have $c$ to be purely real.
After the imposition of the above conditions eq (27) and (28), the Unitary Matrix reduces to:\\
\begin{equation*}
U =
\begin{bmatrix}
    \sqrt{2}\tilde{d} & -\sqrt{2}c \\
    -\sqrt{2}\tilde{c} & -\sqrt{2}d      \\
          
\end{bmatrix}
= \begin{bmatrix}
    -\sqrt{2}d & -\sqrt{2}c \\
    -\sqrt{2}c & -\sqrt{2}d      \\
   
   \end{bmatrix}
\end{equation*}
Where we made use of eq (27) and eq (28) above, $d + \tilde{d} = 0$ or $d = -\tilde{d}$ and $c - \tilde{c} = 0$ or $c = \tilde{c}$
\\
\\
det(U) = $2d^2 - 2c^2 = -1$ or $2c^2 - 2d^2 = 1$\\
\\
Using the above matrix, Alice successfully retrieves the state as follows:
\[
\begin{bmatrix}
    -\sqrt{2}d & -\sqrt{2}c \\
    -\sqrt{2}c & -\sqrt{2}d      \\
\end{bmatrix}
\begin{bmatrix}
    \sqrt{2}(d\alpha - c\beta) \\
    -\sqrt{2}(c\alpha - d\beta)    \\
   
\end{bmatrix}
= 
\begin{bmatrix}
    2(-d^2\alpha + cd\beta + c^2\alpha - cd\beta)      \\
    2(-cd\alpha + c^2\beta + cd\alpha - d^2\beta )    
\end{bmatrix} 
=
\begin{bmatrix}
    2(c^2 - d^2)\alpha      \\
    2(c^2 - d^2)\beta      
\end{bmatrix} 
=
\begin{bmatrix}
    \alpha      \\
    \beta      
\end{bmatrix} 
\]
Expressing the above matrix in terms of Pauli Matrices, we have U = $-\sqrt{2}(dI + c\sigma_x)$
\\
\\
Now as $2c^2 - 2d^2 = 1$, we have $d^2 = c^2 - \frac{1}{2}$
\\
\\
Since $d$ is purely imaginary, we must have $c^2 \leq \frac{1}{2}$ or $-\frac{1}{\sqrt{2}} \leq c \leq \frac{1}{\sqrt{2}}$
Now from the condition of normalization of Alice's resultant state, we have:
\begin{eqnarray}
|d|^2+|c|^2+(\tilde{d}c+d\tilde{c})(\alpha\tilde\beta+\tilde\alpha\beta)
 =  \frac{1}{2} {}\nonumber&&  
\end{eqnarray}
As $c = \tilde{c}$, the above equation reduces to: 
\begin{eqnarray}
|d|^2+|c|^2+c(\tilde{d}+d)(\alpha\tilde\beta+\tilde\alpha\beta)
 =  \frac{1}{2} {}\nonumber   
\end{eqnarray}
And as $\tilde{d}+d$ = 0, the above equation further reduces to:
\begin{eqnarray}
|d|^2+|c|^2 = \frac{1}{2} {}\nonumber&&  
\end{eqnarray}
As $c$ is a real number, $c^2 = |c|^2$ and so we have:
\begin{eqnarray}
|d|^2+c^2 =  \frac{1}{2}{}\nonumber &&  
\end{eqnarray}
or
\begin{eqnarray}
|d|^2 =  \frac{1}{2} - c^2 {}\nonumber&&  
\end{eqnarray}
\\

\noindent\textit{\bf Appendix 4:} (Proof of Theorem 1)\\ \\
For Alice to successfully retrieve the state, Bob and Charlie must measure their respective qubits in the Hadamard Basis. We show how Alice can retrieve back her message by enlisting down the respective local operations corresponding to Bob’s and Charlie’s measurement outcomes.\\ \\
If Bob measures his qubit in Hadamard Basis $\lvert+\rangle$ and Charlie measures his respective qubit in Hadamard Basis $\lvert+\rangle$, the resultant state on Alice’s side is given by the following (when Alice's initial measurement outcome is $\lvert\psi^+\rangle$):
\begin{eqnarray}
\sqrt{2}\{(b\alpha + a\beta)|0\rangle + (a\alpha + b\beta)|1\rangle\} {}\nonumber
\end{eqnarray}
Condition of normalization of the above Alice's resultant state is as follows:
\begin{eqnarray}
|a|^2+|b|^2+(\tilde{a}b+a\tilde{b})(\alpha\tilde\beta+\tilde\alpha\beta)
 =  \frac{1}{2} {}\nonumber&&  
\end{eqnarray}
In order to find Alice's Local Operations state for retrieving the state, we proceed by taking a Unitary Matrix as follows:
\begin{equation*}
U = 
\begin{bmatrix}
a_{1} + ib_{1} & c_{1} - id_{1} \\
c_{1} + id_{1} & -a_{1} + ib_{1} \\
\end{bmatrix}
\end{equation*}
such that det(U) = -1
\\
\\
We need to find $a_{1}, b_{1}, c_{1}, d_{1}$ by the solving the following equation for Alice to retrieve the state:

\[
\begin{bmatrix}
    a_{1} + ib_{1} & c_{1} - id_{1} \\
    c_{1} + id_{1} & -a_{1} + ib_{1}      \\
          
\end{bmatrix}
\begin{bmatrix}
    \sqrt{2}(b\alpha + a\beta) \\
    \sqrt{2}(a\alpha + b\beta)    \\
          
\end{bmatrix}
= 
\begin{bmatrix}
    \alpha      \\
    \beta      
\end{bmatrix} 
\]
\\
As $a, b$, $\alpha$ and $\beta$ are complex numbers, we write $a = x_{1} + iy_{1}, b = x_{2} + iy_{2}, \alpha = x_{3} + iy_{3}, \beta = x_{4} + iy_{4}$. 
\\
The above equation transforms as follows:
\[
\begin{bmatrix}
    a_{1} + ib_{1} & c_{1} - id_{1} \\
    c_{1} + id_{1} & -a_{1} + ib_{1}      \\
          
\end{bmatrix}
\begin{bmatrix}
    \sqrt{2}(\lambda_{1} + i\lambda_{2}) \\
    \sqrt{2}(\lambda_{3} + i\lambda_{4})    \\
          
\end{bmatrix}
= 
\begin{bmatrix}
    \alpha      \\
    \beta      
\end{bmatrix} 
\]
\\
Where we have:
\begin{eqnarray}
\lambda_{1} = x_{2}x_{3} + x_{1}x_{4} - y_{1}y_{4} - y_{2}y_{3}{}\nonumber\\
\lambda_{2} = x_{3}y_{2} + x_{4}y_{1} + x_{2}y_{3} + x_{1}y_{4}{}\nonumber\\
\lambda_{3} = x_{1}x_{3} + x_{2}x_{4} - y_{1}y_{3} - y_{2}y_{4}{}\nonumber\\
\lambda_{4} = x_{3}y_{1} + x_{1}y_{3} + x_{2}y_{4} + x_{4}y_{2}{}\nonumber
\end{eqnarray}
\\
Such that:
\begin{eqnarray}
\lambda_{1}^2 + \lambda_{2}^2 + \lambda_{3}^2 + \lambda_{4}^2 = |a|^2+|b|^2+(\tilde{a}b+a\tilde{b})(\alpha\tilde\beta+\tilde\alpha\beta) =  \frac{1}{2}
\end{eqnarray}
Solving the above equation by comparing real and imaginary parts on both sides to get:
\begin{eqnarray}
a_{1} = \frac{\lambda_{1}x_{3} + \lambda_{2}y_{3} - \lambda_{3}x_{4} - \lambda_{4}y{4}}{\sqrt{2}{(\lambda_{1}^2 + \lambda_{2}^2 + \lambda_{3}^2 + \lambda_{4}^2)}} = \sqrt{2}(\lambda_{1}x_{3} + \lambda_{2}y_{3} - \lambda_{3}x_{4} - \lambda_{4}y_{4}) {}\nonumber\\
b_{1} = \frac{\lambda_{1}y_{3} - \lambda_{2}x_{3} + \lambda_{3}y_{4} - \lambda_{4}x_{4}}{\sqrt{2}{(\lambda_{1}^2 + \lambda_{2}^2 + \lambda_{3}^2 + \lambda_{4}^2)}} = \sqrt{2}(\lambda_{1}y_{3} - \lambda_{2}x_{3} + \lambda_{3}y_{4} - \lambda_{4}x_{4}) {}\nonumber\\
c_{1} = \frac{\lambda_{1}x_{4} + \lambda_{2}y_{4} + \lambda_{3}x_{3} + \lambda_{4}y_{3}}{\sqrt{2}{(\lambda_{1}^2 + \lambda_{2}^2 + \lambda_{3}^2 + \lambda_{4}^2)}} = \sqrt{2}(\lambda_{1}x_{4} + \lambda_{2}y_{4} + \lambda_{3}x_{3} + \lambda_{4}y_{3}) {}\nonumber\\
d_{1} = \frac{\lambda_{4}x_{3} - \lambda_{3}y_{3} + \lambda_{1}y_{4} - \lambda_{2}x_{4}}{\sqrt{2}{(\lambda_{1}^2 + \lambda_{2}^2 + \lambda_{3}^2 + \lambda_{4}^2)}} = \sqrt{2}(\lambda_{4}x_{3} - \lambda_{3}y_{3} + \lambda_{1}y_{4} - \lambda_{2}x_{4}) {}\nonumber
\end{eqnarray}
\begin{eqnarray}
a_{1} + ib_{1} = \sqrt{2}\{(\lambda_{1} - i\lambda_{2})\alpha - (\lambda_{3} + i\lambda_{4})\tilde{\beta}\} {}\nonumber\\
= \sqrt{2}\{(\tilde{b}\tilde{\alpha} + \tilde{a}\tilde{\beta})\alpha - (a\alpha + b\beta)\tilde{\beta}\} {}\nonumber\\
= \sqrt{2}\{(\tilde{b} - (\tilde{b} + b)|\beta|^2 + (\tilde{a} - a)\alpha\tilde{\beta})\} 
\end{eqnarray}
\begin{eqnarray}
-a_{1} + ib_{1} = \sqrt{2}\{-(\lambda_{1} + i\lambda_{2})\tilde{\alpha} + (\lambda_{3} - i\lambda_{4})\beta\} {}\nonumber\\
= \sqrt{2}\{-(b\alpha + a\beta)\tilde{\alpha} + (\tilde{b}\tilde{\beta} + \tilde{a}\tilde{\alpha})\beta\} {}\nonumber\\
= \sqrt{2}\{-b + (\tilde{b} + b)|\beta|^2 + (\tilde{a} - a)\tilde{\alpha}\beta\} 
\end{eqnarray}
\begin{eqnarray}
c_{1} + id_{1} = \sqrt{2}\{(\lambda_{1} - i\lambda_{2})\beta + (\lambda_{3} + i\lambda_{4})\tilde{\alpha}\} {}\nonumber\\
= \sqrt{2}\{(\tilde{b}\tilde{\alpha} + \tilde{a}\tilde{\beta})\beta + (a\alpha + b\beta)\tilde{\alpha}\} {}\nonumber\\
= \sqrt{2}\{\tilde{a} + (\tilde{b} + b)\tilde{\alpha}\beta - (\tilde{a} - a)|\alpha|^2\}
\end{eqnarray}
\begin{eqnarray}
c_{1} - id_{1} = \sqrt{2}\{(\lambda_{1} + i\lambda_{2})\tilde{\beta} + (\lambda_{3} - i\lambda_{4})\alpha\} {}\nonumber\\
= \sqrt{2}\{(b\alpha + a\beta)\tilde{\beta} + (\tilde{b}\tilde{\beta} + \tilde{a}\tilde{\alpha})\alpha\} {}\nonumber\\
= \sqrt{2}\{a + (\tilde{b} + b)\alpha\tilde{\beta} + (\tilde{a} - a)|\alpha|^2\}
\end{eqnarray}
Using equations (30), (31), (32) and (33), we obtain the Unitary matrix as follows:\\
\begin{equation*}
\begin{bmatrix}
    \sqrt{2}\{(\tilde{b} - (\tilde{b} + b)|\beta|^2 + (\tilde{a} - a)\alpha\tilde{\beta})\} & \sqrt{2}\{a + (\tilde{b} + b)\alpha\tilde{\beta} + (\tilde{a} - a)|\alpha|^2\} \\
    \sqrt{2}\{\tilde{a} + (\tilde{b} + b)\tilde{\alpha}\beta - (\tilde{a} - a)|\alpha|^2\} & \sqrt{2}\{-b + (\tilde{b} + b)|\beta|^2 + (\tilde{a} - a)\tilde{\alpha}\beta\}      \\
          
\end{bmatrix}
\end{equation*}
\\
\\
Since the elements of the matrix need to be independent of $\alpha$ and $\beta$, we have the following conditions:\\
\begin{eqnarray}
\tilde{b} + b = 0
\end{eqnarray}
\begin{eqnarray}
\tilde{a} - a = 0
\end{eqnarray}
From eq (34) we have $b$ to be purely imaginary and from eq (35) we have $a$ to be purely real.
After the imposition of the above conditions eq (34) and (35), the Unitary Matrix reduces to:\\
\begin{equation*}
U =
\begin{bmatrix}
    \sqrt{2}\tilde{b} & \sqrt{2}a \\
    \sqrt{2}\tilde{a} & -\sqrt{2}b      \\
   \end{bmatrix}
= \begin{bmatrix}
    -\sqrt{2}b & \sqrt{2}a \\
    \sqrt{2}a & -\sqrt{2}b      \\
  \end{bmatrix}
\end{equation*}
Where we made use of eq (34) and eq (35) above, $\tilde{a} - a = 0$ or $\tilde{a} = a$ and $\tilde{b} + b = 0$ or $\tilde{b} = -b$ 
\\
\\
det(U) = $2b^2 - 2a^2 = -1$ or $2a^2 - 2b^2 = 1$\\
\\
Using the above matrix, Alice successfully retrieves the state as follows:
\[
\begin{bmatrix}
    -\sqrt{2}b & \sqrt{2}a \\
    \sqrt{2}a & -\sqrt{2}b      \\
  \end{bmatrix}
\begin{bmatrix}
    \sqrt{2}(b\alpha + a\beta) \\
    \sqrt{2}(a\alpha + b\beta)    \\
 \end{bmatrix}
= 
\begin{bmatrix}
    2(-b^2\alpha - ab\beta + a^2\alpha + ab\beta)      \\
    2(ab\alpha + a^2\beta -ab\alpha - b^2\beta )    
\end{bmatrix} 
=
\begin{bmatrix}
    2(a^2 - b^2)\alpha      \\
    2(a^2 - b^2)\beta      
\end{bmatrix} 
=
\begin{bmatrix}
    \alpha      \\
    \beta      
\end{bmatrix} 
\]
\\
Expressing the above matrix in terms of Pauli Matrices, we have U = $\sqrt{2}(-bI + a\sigma_x)$
\\
\\
Now as $2a^2 - 2b^2 = 1$, we have $b^2 = a^2 - \frac{1}{2}$
\\
\\
Since $b$ is purely imaginary, we must have $a^2 \leq \frac{1}{2}$ or $-\frac{1}{\sqrt{2}} \leq a \leq \frac{1}{\sqrt{2}}$
Now from the condition of normalization of Alice's resultant state, we have:
\begin{eqnarray}
|a|^2+|b|^2+(\tilde{a}b+a\tilde{b})(\alpha\tilde\beta+\tilde\alpha\beta)
 =  \frac{1}{2} {}\nonumber&&  
\end{eqnarray}
As $a = \tilde{a}$, the above equation reduces to: 
\begin{eqnarray}
|a|^2+|b|^2+a(\tilde{b}+b)(\alpha\tilde\beta+\tilde\alpha\beta)
 =  \frac{1}{2} {}\nonumber   
\end{eqnarray}
And as $\tilde{b}+b$ = 0, the above equation further reduces to:
\begin{eqnarray}
|a|^2+|b|^2 = \frac{1}{2} {}\nonumber&&  
\end{eqnarray}
As $a$ is a real number, $a^2 = |a|^2$ and so we have:
\begin{eqnarray}
|b|^2+a^2 =  \frac{1}{2}{}\nonumber &&  
\end{eqnarray}
or
\begin{eqnarray}
|b|^2 =  \frac{1}{2} - a^2 {}\nonumber&&  
\end{eqnarray}
\\

\noindent\textit{\bf Appendix 5:} (Proof of Theorem 2)\\ \\
For Charlie to successfully reconstruct the state, Alice and Bob must measure their respective qubits in the Hadamard Basis. We show how Charlie can reconstruct the state by enlisting down the respective local operations corresponding to Alice's and Bob’s measurement outcomes. \\
If Alice and Bob’s joint measurement outcome in Hadamard Basis is $\lvert++\rangle$, the resultant state on Charlie’s side is given by the following (when Alice’s initial measurement outcome is $\lvert\phi^+\rangle$):
\begin{eqnarray}
\frac{1}{\sqrt{2}}\{[(a+b-c+d)\alpha+(a+b+c-d)\beta]|0\rangle\}+\frac{1}{\sqrt{2}}\{[(a+b+c-d)\alpha + (a+b-c+d)\beta]|1\rangle\} {}\nonumber&&
\end{eqnarray}
Denoting $(a+b-c+d)$ as $\lambda_1$ and $(a+b+c-d)$ as $\lambda_2$, the above resultant state can be written as:
\begin{eqnarray}
\frac{1}{\sqrt{2}}(\lambda_1\alpha+\lambda_2\beta)|0\rangle + \frac{1}{\sqrt{2}}(\lambda_2\alpha+\lambda_1\beta)|1\rangle {}\nonumber
\end{eqnarray}
Condition of normalization of the above Charlie's resultant state is as follows: 
\begin{eqnarray}
\frac{1}{2}\{\lvert\lambda_1\alpha+\lambda_2\beta\rvert^2 + \lvert\lambda_2\alpha+\lambda_1\beta\rvert^2 \} = 1 {}\nonumber&&
\end{eqnarray}
Simplifying, we get:
\begin{eqnarray}
\lvert\lambda_1\rvert^2+\lvert\lambda_2\rvert^2+(\tilde\lambda_1\lambda_2+\lambda_1\tilde\lambda_2)(\alpha\tilde\beta+\tilde\alpha\beta) = 2 {}\nonumber&&
\end{eqnarray}
\\
In order to find Charlie's Local Operations state for retrieving the state, we proceed by taking a Unitary Matrix as follows:
\begin{equation*}
U = 
\begin{bmatrix}
a_{1} + ib_{1} & c_{1} - id_{1} \\
c_{1} + id_{1} & -a_{1} + ib_{1} \\
\end{bmatrix}
\end{equation*}
such that det(U) = -1
\\
\\
We need to find $a_{1}, b_{1}, c_{1}, d_{1}$ by the solving the following equation for Alice to retrieve the state:
\[
\begin{bmatrix}
    a_{1} + ib_{1} & c_{1} - id_{1} \\
    c_{1} + id_{1} & -a_{1} + ib_{1}      \\
          
\end{bmatrix}
\begin{bmatrix}
    \frac{1}{\sqrt{2}}(\lambda_{1}\alpha + \lambda_{2}\beta) \\
    \frac{1}{\sqrt{2}}(\lambda_{2}\alpha + \lambda_{1}\beta)    \\
          
\end{bmatrix}
= 
\begin{bmatrix}
    \alpha      \\
    \beta      
\end{bmatrix} 
\]
\\
As $a$, $b$, $c$, $d$, $\alpha$ and $\beta$ are complex numbers, we write $\lambda_{1} = x_{1} + iy_{1}, \lambda_{2} = x_{2} + iy_{2}, \alpha = x_{3} + iy_{3}, \beta = x_{4} + iy_{4}$. 
\\
\\
The above equation transforms as follows:
\[
\begin{bmatrix}
    a_{1} + ib_{1} & c_{1} - id_{1} \\
    c_{1} + id_{1} & -a_{1} + ib_{1}      \\
          
\end{bmatrix}
\begin{bmatrix}
    \frac{1}{\sqrt{2}}(\nu_{1} + i\nu_{2}) \\
    \frac{1}{\sqrt{2}}(\nu_{3} + i\nu_{4})    \\
          
\end{bmatrix}
= 
\begin{bmatrix}
    \alpha      \\
    \beta      
\end{bmatrix} 
\]
\\
Where we have:\\
\begin{eqnarray}
\nu_{1} = x_{1}x_{3} + x_{2}x_{4} - y_{1}y_{3} - y_{2}y_{4}{}\nonumber\\
\nu_{2} = x_{3}y_{1} + x_{1}y_{3} + x_{2}y_{4} + x_{4}y_{2}{}\nonumber\\
\nu_{3} = x_{2}x_{3} + x_{1}x_{4} - y_{1}y_{4} - y_{2}y_{3}{}\nonumber\\
\nu_{4} = x_{3}y_{2} + x_{4}y_{1} + x_{2}y_{3} + x_{1}y_{4}{}\nonumber
\end{eqnarray}
\\
Such that:\\
\begin{eqnarray}
\nu_{1}^2 + \nu_{2}^2 + \nu_{3}^2 + \nu_{4}^2 = |\lambda_1|^2+|\lambda_2|^2+(\tilde\lambda_1\lambda_2+\lambda_1\tilde\lambda_2)(\alpha\tilde\beta+\tilde\alpha\beta) =  2
\end{eqnarray}
Solving the above equation by comparing real and imaginary parts on both sides to get:
\begin{eqnarray}
a_{1} = \frac{\sqrt{2}(\nu_{1}x_{3} + \nu_{2}y_{3} - \nu_{3}x_{4} - \nu_{4}y_{4})}{\nu_{1}^2 + \nu_{2}^2 + \nu_{3}^2 + \nu_{4}^2} = \frac{ \nu_{1}x_{3} + \nu_{2}y_{3} - \nu_{3}x_{4} - \nu_{4}y_{4}}{\sqrt{2}} {}\nonumber\\
b_{1} = \frac{\sqrt{2}(\nu_{1}y_{3} - \nu_{2}x_{3} + \nu_{3}y_{4} - \nu_{4}x_{4})}{\nu_{1}^2 + \nu_{2}^2 + \nu_{3}^2 + \nu_{4}^2} = \frac{ \nu_{1}y_{3} - \nu_{2}x_{3} + \nu_{3}y_{4} - \nu_{4}x_{4}}{\sqrt{2}} {}\nonumber\\
c_{1} = \frac{{\sqrt{2}}(\nu_{1}x_{4} + \nu_{2}y_{4} + \nu_{3}x_{3} + \nu_{4}y_{3})}{\nu_{1}^2 + \nu_{2}^2 + \nu_{3}^2 + \nu_{4}^2} = \frac{ \nu_{1}x_{4} + \nu_{2}y_{4} + \nu_{3}x_{3} + \nu_{4}y_{3}}{\sqrt{2}} {}\nonumber\\
d_{1} = \frac{{\sqrt{2}}(\nu_{4}x_{3} - \nu_{3}y_{3} + \nu_{1}y_{4} - \nu_{2}x_{4})}{\nu_{1}^2 + \nu_{2}^2 + \nu_{3}^2 + \nu_{4}^2} = \frac{ \nu_{4}x_{3} - \nu_{3}y_{3} + \nu_{1}y_{4} - \nu_{2}x_{4}}{\sqrt{2}} {}\nonumber
\end{eqnarray}
\begin{eqnarray}
a_{1} + ib_{1} = \frac{1}{\sqrt{2}}\{(\nu_{1} - i\nu_{2})\alpha - (\nu_{3} + i\nu_{4})\tilde{\beta}\} {}\nonumber\\
= \frac{1}{\sqrt{2}}\{(\tilde{\lambda_1}\tilde{\alpha} + \tilde{\lambda_1}\tilde{\beta})\alpha - (\lambda_2\alpha + \lambda_1\beta)\tilde{\beta}\} {}\nonumber\\
= \frac{1}{\sqrt{2}}\{(\tilde{\lambda_1} - (\tilde{\lambda_1} + \lambda_1)|\beta|^2 + (\tilde{\lambda_2} - \lambda_2)\alpha\tilde{\beta})\} 
\end{eqnarray}
\begin{eqnarray}
-a_{1} + ib_{1} = \frac{1}{\sqrt{2}}\{-(\nu_{1} + i\nu_{2})\tilde{\alpha} + (\nu_{3} - i\nu_{4})\beta\} {}\nonumber\\
= \frac{1}{\sqrt{2}}\{-(\lambda_1\alpha + \lambda_2\beta)\tilde{\alpha} + (\tilde{\lambda_1}\tilde{\beta} + \tilde{\lambda_2}\tilde{\alpha})\beta\} {}\nonumber\\
= \frac{1}{\sqrt{2}}\{-\lambda_1 + (\tilde{\lambda_1} + \lambda_1)|\beta|^2 + (\tilde{\lambda_2} - \lambda_2)\tilde{\alpha}\beta\} 
\end{eqnarray}
\begin{eqnarray}
c_{1} + id_{1} = \frac{1}{\sqrt{2}}\{(\nu_{1} - i\nu_{2})\beta + (\nu_{3} + i\nu_{4})\tilde{\alpha}\} {}\nonumber\\
= \frac{1}{\sqrt{2}}\{(\tilde{\lambda_1}\tilde{\alpha} + \tilde{\lambda_2}\tilde{\beta})\beta + (\lambda_2\alpha + \lambda_1\beta)\tilde{\alpha}\} {}\nonumber\\
= \frac{1}{\sqrt{2}}\{\tilde{\lambda_2} + (\tilde{\lambda_1} + \lambda_1)\tilde{\alpha}\beta - (\tilde{\lambda_2} - \lambda_2)|\alpha|^2\}
\end{eqnarray}
\begin{eqnarray}
c_{1} - id_{1} = \frac{1}{\sqrt{2}}\{(\nu_{1} + i\nu_{2})\tilde{\beta} + (\nu_{3} - i\nu_{4})\alpha\} {}\nonumber\\
= \frac{1}{\sqrt{2}}\{(\lambda_1\alpha + \lambda_2\beta)\tilde{\beta} + (\tilde{\lambda_1}\tilde{\beta} + \tilde{\lambda_2}\tilde{\alpha})\alpha\} {}\nonumber\\
= \frac{1}{\sqrt{2}}\{\lambda_2 + (\tilde{\lambda_1} + \lambda_1)\alpha\tilde{\beta} + (\tilde{\lambda_2} - \lambda_2)|\alpha|^2\}
\end{eqnarray}
Using equations (38), (39), (40) and (41), we obtain the Unitary matrix as follows:\\
\begin{equation*}
\begin{bmatrix}
    \frac{1}{\sqrt{2}}\{(\tilde{\lambda_1} - (\tilde{\lambda_1} + \lambda_1)|\beta|^2 + (\tilde{\lambda_2} - \lambda_2)\alpha\tilde{\beta})\} & \frac{1}{\sqrt{2}}\{\lambda_2 + (\tilde{\lambda_1} + \lambda_1)\alpha\tilde{\beta} + (\tilde{\lambda_2} - \lambda_2)|\alpha|^2\} \\
    \frac{1}{\sqrt{2}}\{\tilde{\lambda_2} + (\tilde{\lambda_1} + \lambda_1)\tilde{\alpha}\beta - (\tilde{\lambda_2} - \lambda_2)|\alpha|^2\} & \frac{1}{\sqrt{2}}\{-\lambda_1 + (\tilde{\lambda_1} + \lambda_1)|\beta|^2 + (\tilde{\lambda_2} - \lambda_2)\tilde{\alpha}\beta\}      \\
          
\end{bmatrix}
\end{equation*}
\\
\\
Since the elements of the matrix need to be independent of $\alpha$ and $\beta$, we have the following conditions:\\
\begin{eqnarray}
\tilde{\lambda_1} + \lambda_1 = 0
\end{eqnarray}
\begin{eqnarray}
\tilde{\lambda_2} - \lambda_2 = 0
\end{eqnarray}
From eq (42) we have $\lambda_1$ or $(a+b-c+d)$ to be purely imaginary and from eq (43) we have $\lambda_2$ or $(a+b+c-d)$ to be purely real.
After the imposition of the above conditions eq (42) and (43), the Unitary Matrix reduces to:\\
\begin{equation*}
U =
\begin{bmatrix}
    \frac{1}{\sqrt{2}}\tilde{\lambda_1} & \frac{1}{\sqrt{2}}\lambda_2 \\
    \frac{1}{\sqrt{2}}\tilde{\lambda_2} & -\frac{1}{\sqrt{2}}\lambda_1      \\
          
\end{bmatrix}
= \begin{bmatrix}
    -\frac{1}{\sqrt{2}}\lambda_1 & \frac{1}{\sqrt{2}}\lambda_2 \\
    \frac{1}{\sqrt{2}}\lambda_2 & -\frac{1}{\sqrt{2}}\lambda_1      \\
          
\end{bmatrix}
\end{equation*}
Where we made use of eq (41) and eq (42) above, $\tilde{\lambda_1} + \lambda_1 = 0$ or $\tilde{\lambda_1} = -\lambda_1$ and $\tilde{\lambda_2} - \lambda_2 = 0$ or $\tilde{\lambda_2} = \lambda_2$
\\
\\
det(U) = $\frac{1}{2}\lambda_1^2 - \frac{1}{2}\lambda_2^2 = -1$ or $\frac{1}{2}\lambda_2^2 - \frac{1}{2}\lambda_1^2 = 1$\\
\\
Using the above matrix, Alice successfully retrieves the state as follows:
\[
\begin{bmatrix}
    -\frac{1}{\sqrt{2}}\lambda_1 & \frac{1}{\sqrt{2}}\lambda_2 \\
    \frac{1}{\sqrt{2}}\lambda_2 & -\frac{1}{\sqrt{2}}\lambda_1      \\
          
\end{bmatrix}
\begin{bmatrix}
    \frac{1}{\sqrt{2}}(\lambda_1\alpha + \lambda_2\beta) \\
    \frac{1}{\sqrt{2}}(\lambda_2\alpha + \lambda_1\beta)    \\
          
\end{bmatrix}
= 
\begin{bmatrix}
    \frac{1}{2}(-\lambda_1^2\alpha - \lambda_1\lambda_2\beta + \lambda_2^2\alpha + \lambda_1\lambda_2\beta)      \\
    \frac{1}{2}(\lambda_1\lambda_2\alpha + \lambda_2^2\beta -\lambda_1\lambda_2\alpha - \lambda_1^2\beta )    
\end{bmatrix} 
=
\begin{bmatrix}
    \frac{1}{2}(\lambda_2^2 - \lambda_1^2)\alpha      \\
    \frac{1}{2}(\lambda_2^2 - \lambda_1^2)\beta      
\end{bmatrix} 
=
\begin{bmatrix}
    \alpha      \\
    \beta      
\end{bmatrix} 
\]
Expressing the above matrix in terms of Pauli Matrices, we have U = $\frac{1}{\sqrt{2}}(-\lambda_1I + \lambda_2\sigma_x)$ or U = $\frac{1}{\sqrt{2}}[-(a+b-c+d)I + (a+b+c-d)\sigma_x]$
\\ \\
From eq (41) we derived $\lambda_1$ or $(a+b-c+d)$ to be purely imaginary and from eq (42) we derived $\lambda_2$ or $(a+b+c-d)$ to be purely real, so we have the following conditions of $a$, $b$, $c$ and $d$ from Alice's retrieval derived in Appendix 1:\\
\\
\textbf{Case:} $(a+b-c+d)$ is purely imaginary and $(a+b+c-d)$ is purely real\\
If $a$ and $c$ are chosen as real numbers, and $b$ and $d$ are chosen as imaginary numbers, then we have $a$ = $c$ and $b = d$\\
If $a$ and $d$ are chosen as real numbers, and $b$ and $c$ are chosen as imaginary numbers, then we have $a$ = $-d$ and $b = -c$\\
If $b$ and $c$ are chosen as real numbers, and $a$ and $d$ are chosen as imaginary numbers, then we have $b$ = $c$ and $a = d$\\
If $b$ and $d$ are chosen as real numbers, and $a$ and $c$ are chosen as imaginary numbers, then we have $b$ = $-d$ and $a = -c$\\
\\
\end{widetext}
\end{document}